# Identifying Contact Barrier Types in Few-Layer $MoS_2$ Devices Using Correlative IV, LBIC, and Bias-Dependent KPFM

Ariane Ufer[1,2], Zeinab Eftekhari[3], Benjamin Mayer[1], Hendrik Lambers[1,2], Hubert J. Krenner[1], Rebecca Saive[3,*], Ursula Wurstbauer[1,2,*]

[1] Institute of Physics, University of Münster, Münster, Germany

[2] Center for Soft Nanoscience (SoN), University of Münster, Münster, Germany

[3] MESA+ Institute for Nanotechnology, University of Twente, Enschede, The Netherlands

*Corresponding authors: wurstbauer@uni-muenster.de; r.saive@utwente.nl

**Abstract**

Electrical contacts between metals and two-dimensional (2D) semiconductors such as molybdenum disulfide ($MoS_2$) critically govern device performance, yet their microscopic nature remains difficult to disentangle using any single characterization technique. Here we present an integrated experimental framework that combines current–voltage (IV) characterization, laser beam induced current (LBIC) mapping, and bias-dependent Kelvin probe force microscopy (KPFM) to comprehensively resolve the contact properties of few-layer $MoS_2$-based two-terminal devices under ambient conditions. IV measurements deliver macroscopic transport characteristics as a function of bias voltage and illumination conditions. LBIC maps the local photocurrent response with micrometer spatial resolution, revealing the position and nature of internal electric fields at $MoS_2$–metal interfaces. KPFM, operated under an applied static bias rather than in the conventional work-function mode, provides nanoscale-resolved potential distributions that quantify the relative magnitudes and spatial locations of contact barriers. We apply this framework to three representative devices — one exhibiting ohmic-like and two exhibiting diode-like contact behavior — and demonstrate that the combined analysis can unambiguously identify whether the dominant barrier is of Schottky or tunnel type and determine the asymmetry between the two contacts. We further demonstrate that thermal annealing significantly reduces the total resistance, while contact barriers remain the dominant source of resistance. The methodology is directly transferable to other 2D semiconductor–metal systems and provides a practical yet comprehensive route toward a quantitative microscopic understanding of 2D device contacts.

## 1. Introduction

Two-dimensional (2D) semiconductors, and in particular transition metal dichalcogenides (TMDCs) such as molybdenum disulfide ($MoS_2$), have attracted sustained research interest over the past 15 years owing to their exceptional and thickness-tunable electronic and optical properties, including direct band gaps in the monolayer limit, strong light–matter coupling, and mechanical flexibility that is unattainable in bulk semiconductors.[1–3] These characteristics position $MoS_2$ and related TMDCs as compelling channel materials for next-generation field-effect transistors (FETs), photodetectors, and photovoltaic devices.[4,5] Translating these intrinsic material properties into high-performance devices, however, critically depends on the quality of the electrical contacts between the 2D semiconductor and the metal electrodes — a challenge that has proven to be more complex than conventional bulk-semiconductor counterparts.[6–8]

Unlike conventional bulk semiconductors, 2D materials are free of surface dangling bonds and form van der Waals (vdW) interfaces with deposited metals, giving rise to a rich and often counterintuitive contact phenomenology.[1,6,9,10] Schottky barriers, Fermi-level pinning, and tunnel barriers can all be present simultaneously, and their relative contributions depend sensitively on parameters such as flake thickness, contact metal work function, interface cleanliness, fabrication route, and post-processing treatments.[6,7,11–16] Fermi-level pinning near the conduction band edge of $MoS_2$, for instance, is now understood to be nearly universal across a wide range of contact metals,[11,17] yet the degree of pinning and its consequences for device resistance vary enormously between samples.[10,17] This near-universal pinning makes the absolute barrier height difficult to predict from metal work functions alone, motivating the need for device-level characterization. Layer thickness plays a particularly decisive role: five-layer $MoS_2$ has been identified as an optimal thickness for minimizing contact resistance,[10] while mono- and bilayer devices are dominated by substantially larger barriers.[7,10,12]

Extensive experimental effort has been devoted to characterize $MoS_2$–metal contacts using a variety of individual techniques, including temperature-dependent IV measurements, transfer-length methods, photoemission spectroscopy, acousto-electric transport measurements and scanning probe approaches.[6–8,11,13–28] Kelvin probe force microscopy (KPFM) has been employed to map work functions and surface potentials at 2D–metal interfaces, providing insight into band alignment and Fermi-level pinning.[12,23,24,29] Scanning photocurrent measurements, also known as laser beam induced current (LBIC) mapping, have revealed the spatial distribution of internal electric fields and identified the dominant photocurrent generation mechanisms (photovoltaic, photothermoelectric (PTE), and photoconductive) at individual junctions.[5,9,30] Despite these efforts, a comprehensive picture of the interplay between Schottky barriers, tunnel barriers, and Fermi-level pinning at individual $MoS_2$–metal junctions remains elusive, in part because no single technique captures both the macroscopic transport behavior and the microscopic spatial distribution of barriers within a single device.[5,6,14]

Contact optimization strategies, including the selection of low-work-function metals, edge-contact geometries, surface doping, strain engineering, and thermal annealing have yielded significant improvements in individual devices, yet a predictive understanding of which barrier type dominates in a given sample, and how it responds to a given treatment, remains largely lacking.[6–8,16,18,24,28,29,31–35] This gap is particularly consequential for the rational design of low-resistance ohmic contacts, which is widely recognized as one of the key bottlenecks preventing 2D semiconductor devices from reaching their theoretical performance limits.[6,14,36,37]

Addressing this need requires a combination of complementary experimental approaches that captures both the macroscopic transport behavior and the microscopic spatial distribution of barriers within a single device, without requiring cryogenic conditions, ultra-high vacuum, or gate electrodes. In this work, we present such a framework, combining IV characterization, LBIC mapping, and bias-dependent KPFM measurements performed entirely under ambient conditions with and without illumination. The key methodological advance over prior work lies in the use of KPFM under illumination and an applied static bias voltage[38], rather than in the conventional zero-bias work-function mapping mode[5,7,23,24,29]. Under an applied static bias voltage, KPFM directly maps the spatially resolved voltage drop across the device and thereby quantifies the relative sizes of individual contact barriers without assumptions about the band structure.

We demonstrate the framework on three few-layer $MoS_2$ devices: one five-layer sample with ohmic-like contact behavior (O1), one bilayer sample with diode-like behavior (D1), and one bilayer sample with diode-like behavior studied before and after thermal annealing (D2). By analyzing the results of all three methods in combination, we can (i) determine whether the dominant contact barrier is of Schottky or tunnel type, (ii) identify the relative magnitudes of the barriers at the two contacts, and (iii) track how thermal annealing modifies the individual barrier contributions. The methodology is directly applicable to other 2D semiconductor–metal systems, offering a practical route toward the quantitative microscopic understanding needed to guide rational contact engineering in 2D devices.

## 2. $MoS_2$ Devices and Measurement Methods

The investigated devices consist of a Si/$SiO_2$ or glass substrate with two physical vapor-deposited Cr/Au contacts. Few-layer $MoS_2$ flakes, micromechanically exfoliated from bulk crystals, are transferred onto the contacts by a dry viscoelastic stamping technique.[33] A schematic of the device structure is shown in Figure 1. The layer thickness of the flakes is characterized by Raman spectroscopy.[34] Layer thickness strongly influences the contact resistance: five-layer $MoS_2$ provides near-optimal contact resistance, whereas mono- and bilayer flakes exhibit substantially larger barriers.[7,10–12] We therefore fabricated and studied one sample with a five-layer $MoS_2$ flake (ohmic-like behavior, O1) and two bilayer flakes (diode-like behavior, D1 and D2). Figure 1 summarizes the three complementary measurement methods. The chosen geometry with the $MoS_2$ flake transferred on top of the contacts allows simultaneous KPFM access to the flake region between the contacts and on top of them.

All measurements are performed under ambient conditions. IV measurements provide a macroscopic overview of transport properties. IV curves are recorded in the dark and under global laser illumination ($E_{Laser}$ = 2.33 eV, above the $MoS_2$ band gap) at laser powers of 20, 40, 60, 80, and 100 µW. The defocused illumination does not allow for an accurate estimation of the power density. Assuming a laser spot size of roughly 50 µm in diameter, the resulting power density for 100 µW laser power excitation is in the range of $I \approx 5\ {}^{W}/_{cm^2}$. LBIC measurements reveal the local photocurrent response on a micrometer scale. A focused laser spot ($E_{Laser}$ = 2.33 eV, P = 100 µW, diameter ~1 µm, power density of $I < 1 \times 10^4\ {}^{W}/_{cm^2}$) is scanned across the sample in 2 µm steps; the photocurrent and reflected laser intensity are recorded simultaneously at each position.[5,35,39] Maps with and without applied bias voltage were recorded; maps at non-zero bias are corrected for the dark current.

KPFM is performed in frequency-modulation sideband mode. The tip scans the sample and records the contact potential difference (CPD) at each position. A static bias is applied to the sample with one contact set to ground (always the right contact in the presented micrographs associated with KPFM). Each scan line is measured twice: in the forward direction (with the applied bias) and in the backward direction at 0 V.[38] Subtracting the 0 V CPD data from the biased CPD data yields the effective potential distribution induced by the applied voltage, removing contributions from the sample work function and surface effects. In this way the bias-induced surface voltage profile is recorded. Optionally, the $MoS_2$ flake can be globally illuminated during KPFM ($E_{Laser}$ = 1.95 eV, P = 5 mW), with photon energy above the direct optical band gap of bilayer $MoS_2$ (~1.8–1.9 eV) given by the electronic band gap reduced by the exciton binding energy.[2] The lateral resolution of sideband KPFM is ultimately limited by the tip apex radius (~25 nm); we would like to stress that beyond the scope of the current work sub-10 nm resolution has been demonstrated under optimized UHV conditions.[40,41] An overview of the investigated devices and schemata of the setup are provided in the supplemental materials.

## 3. Results

### 3.1 IV Characteristics

Figure 2a,b shows the IV curves of samples O1 and D1 in the dark and under illumination. Sample O1 exhibits linear, symmetric (ohmic) characteristics independent of illumination, with a resistance at $U_{bias}$ = ±0.2 V of $R_{O1}$(±0.2 V) = 0.3 MΩ reflecting highly symmetric, low-resistance contacts. Illumination causes only a marginal current increase, consistent with low-barrier, highly transparent contacts. Sample D1, by contrast, shows strongly asymmetric, saturating IV characteristics with $R_{D1}$(+0.2 V) = 32.8 MΩ and $R_{D1}$(−0.2 V) = 57.1 MΩ in the dark. Illumination increases the current by more than a factor of 40 at an illumination power of 100 µW and modifies the asymmetry ratio, indicating the presence of significant asymmetric contact barriers that are sensitive to photogenerated carriers (see also supplemental materials for further details).

### 3.2 LBIC Mapping

LBIC maps at 0 V bias are shown in Figure 2c,d. For sample O1, symmetric positive and negative local beam induced photocurrent signals appear mainly on the $MoS_2$ flake above the left and right contact directly next to the contact edge. The positions of the photocurrent signals, located above the metallic contact pads rather than in the gap between them, are characteristic of the photothermoelectric (PTE) effect as the dominant photocurrent mechanism. In the PTE effect a local temperature gradient at the metal–semiconductor interface under focused illumination drives a thermoelectric current via the Seebeck effect. This temperature gradient is strongest directly at the contact edge, where the high thermal conductivity of the metal creates a significant laser induced temperature gradient with the $MoS_2$ flake.[5,9] A photovoltaic contribution, which would arise from a space charge region (SCR) generating an in-plane electric field extending into $MoS_2$ in the gap, is not observed indicating low Schottky barriers. The overall LBIC results observed for sample O1 are consistent with symmetric, low-barrier, in agreement with the KPFM findings (Figure 3 a,b) as discussed in detail below.

For sample D1, a negative photocurrent signal is observed predominantly on the flake over the left half of the gap. This spatial distribution is a hallmark of the photovoltaic effect driven by an internal electric field induced by the SCR at the left contact, and indicating the presence of a Schottky barrier that is presumably in coexistence with a less significant tunnel barrier due to a vdW gap at the $MoS_2$ - gold interface.[5,9,30] The asymmetry of the LBIC signal with respect to left and right contacts mirrors the asymmetry of the IV characteristics and is consistent with the findings in KPFM (Figure 3c,d) as discussed below. It is worth mentioning that the total photocurrent magnitudes for sample O1 are notably higher than for D1, consistent with the order-of-magnitude difference in total resistance deduced from IV measurements.

LBIC maps taken at ±0.1 V bias (Figure 3e–h) further demonstrate the polarity-dependence of the local photocurrent pattern. The dominant photocurrent contribution for sample O1 (Figure 3 e,f) is under the contact region and does not change with polarity of the bias voltage consistent with Seebeck-driven photothermoelectric effect (PTE).[5,9] A weaker polarity dependent contribution exist along the whole flake in the channel region between the contact and is assigned to the photoconductive effect due to photo-generated electron-hole pairs effectively enhancing the free charge carrier density and consequently enhancing the channel conductivity under illumination.[5] Bias dependent LBIC maps taken for sample D1 (Figure 3g,h) show a polarity dependent maximum of the photocurrent inside the channel region that is closer to the left contact for negative bias and rather in the center for positive bias with the right contact grounded. This characteristic asymmetric pattern caused by internal fields induced by SCR points towards asymmetric Schottky-contacts with larger SCR on the left compared to the right contact as already indicated by the diode like IV-behavior (further measurements and schematic of the device structure are provided in the supplemental materials).

We note that LBIC measurements alone cannot unequivocally distinguish between Schottky-type barrier and a tunnel barrier and particularly cannot rule-out the existence of a tunnel barrier introduced by the van der Waals gap between the metal contact and $MoS_2$. However the existence of an internal field induced by the SCR can be unambiguously identified by the position of the maximum of the photocurrent amplitude in the gap together with the polarity dependence of the LBIC maps. The LBIC and IV characteristics together suggest the existence of a Schottky barrier. The rather high contact resistance likely points toward an additional tunnel barrier contributing to the overall LBIC and IV signals; their individual contributions are qualitatively resolved by the complementary KPFM analysis below.

### 3.3 Bias-Dependent KPFM

Bias-induced surface voltage profile lines extracted from KPFM measurements at $U_{bias} = \pm 0.3$ V for sample O1 and at $U_{bias} = \pm 0.5$ V for sample D1 are shown in Figure 3a–d. Sample O1 shows a nearly linear potential drop across the flake, independent of voltage polarity and illumination, confirming symmetric ohmic contacts with negligible barriers. Sample D1 exhibits abrupt, asymmetric potential drops concentrated at the contact edges, strongly dependent on voltage polarity. While for positive bias voltages the drop at the left and right contact have similar magnitude, there is a significant polarity dependence. For negative bias voltages, the majority of the bias drops at the left contact. This polarity dependence at the left contact is characteristic of a Schottky barrier that is alternately reverse- and forward-biased.[13] The smaller, polarity-independent component at the right contact is attributed to a tunnel barrier again consistent with the absence of a SCR induced photovoltaic photocurrent in LBIC signals for zero bias voltage. Under illumination the relative voltage drop observed in KPFM at the left contact is consistently increased for positive and negative bias while it is reduced for the right contact. In addition, there is an increased voltage drop across the $MoS_2$ channel under illumination compared to the dark measurements for positive voltage (Figure 3 c) and a reduction for negative bias (Figure 3 d). Light exposure modifies the potential distribution, which is experienced differently by electrons and holes, resulting in the observed complex, polarity-dependent asymmetric behavior. Simultaneously, light exposure impacts the conductivity due to photo-induced free charge carriers reducing the overall resistivity of the material both affecting the transport across the contacts and the $MoS_2$ channel e.g. by photo-induced activation of carriers (photoconductivity), photodoping and photogating effects.[42]

We note that for sample O1 and D1 the apparent lateral gap size moderately differs by about 1 μm between LBIC and KPFM due to the Cr underlayer of the contacts that is not fully covered by gold given the utilized metal evaporation using a shadow mask: Cr is nearly invisible in optical reflectance conducted simultaneously with LBIC to identify the contact edges but detectable by its work function in KPFM, leading to a smaller apparent gap and an additional thin feature in the KPFM data. The contact pads used for the preparation of sample D2 have been prepared by optical lithography and lift-off after metal evaporation such that the lateral gap agrees between LBIC and KPFM.

### 3.4 Combined Analysis: Identifying Barrier Types and Asymmetry

The three methods provide complementary and mutually consistent information. IV measurements yield absolute resistance values and their voltage and illumination dependence. LBIC identifies the location and nature of internal electric fields and is therefore sensitive to photovoltaic contributions due to Schottky barrier induced SCR and PTE contributions. KPFM maps the relative voltage drops across the two contacts under applied bias, directly quantifying barrier asymmetry. The polarity dependence in KPFM in combination with the findings from LBIC and IV investigations allows one to additionally identify the dominant type of the barrier at the contact being either of Schottky- or tunnel-type. Together, for sample O1, all three methods consistently indicate symmetric, low-barrier ohmic contacts. For sample D1, the combined analysis reveals a dominant Schottky barrier at the left contact and a comparatively smaller tunnel barrier at the right contact as dominant contributions to the total resistance across the two-terminal device.

Combining the KPFM-derived proportional voltage drops with the absolute resistance from IV measurements allows rough estimates of the individual contact resistances (see schematic for serial resistor circuitry provided in the supplemental materials). The two-terminal $MoS_2$ devices are modeled as a three-element series circuit $R_{total} = R_{lc} + R_{sh} + R_{rc}$ with $R_{l(r)c}$ the left (right) vertical contact resistance that includes also the $MoS_2$ sheet resistance on top of the contacts and $R_{sh}$ the $MoS_2$ sheet resistance in the gap (channel). The proportional resistance of each element is $R_{prop} = \alpha \times R_{total}$, where α is the fractional voltage drop at that element read from bias induced surface voltage profile from KPFM, and $R_{total}$ is the total device resistance at the KPFM bias voltage deduced from I-V measurements. These estimates, summarized in Table 1, should be treated as order-of-magnitude values, as the different measurement conditions (illumination type, power density, sweep vs. static bias) preclude exact quantitative comparison. At this stage we note that the applied methodology enables a qualitative identification of the barrier types limiting charge injection at the metal–$MoS_2$ junctions and a proportional decomposition of the individual resistance contributions in the two-terminal device also under illumination, which is directly relevant for optoelectronic applications. A quantitative extraction of the Schottky barrier height could be achieved through thermal activation analysis in temperature-dependent IV measurements.[20] Fermi-level pinning and mid-gap contributions could be further addressed in time-resolved and temperature dependent LBIC investigations under finite bias voltage, which provides access to the electric-field dependence of trap-assisted tunnelling through mid-gap states.[43,44]

**Table 1.** Equivalent circuit model parameters for $MoS_2$–metal devices. Summary of proportional voltage drops (α) and proportional resistances ($R_{prop} = \alpha \times R_{total}$) derived from combined KPFM and IV measurements for sample O1 and D1 dark and under global illumination.

| Circuit Element | Barrier Type | O1 (5L $MoS_2$, dark/bright, ±0.3V) | | D1 (2L $MoS_2$, Diode, dark +0.5 V / -0.5V) | | D1 (2L $MoS_2$, Diode, bright +0.5 V / -0.5V) | |
|---|---|---|---|---|---|---|---|
| | | α (%) | $R_{prop}$ (MΩ) | α (%) | $R_{prop}$ (MΩ) | α (%) | $R_{prop}$ (MΩ) |
| $R_{lc}$ (left contact resistance) | *Schottky / Tunnel* | ~23 | ~0.07 | ~38 / ~72 | ~12.58 / ~80.00 | ~48 / ~84 | ~1.8 / ~1.41 |
| $R_{sh}$ (channel resistance) | *Ohmic ($MoS_2$ gap)* | ~57 | ~0.17 | ~8 / ~14 | ~2.65 / ~15.56 | ~22 / ~8 | ~0.83 / ~0.13 |
| $R_{rc}$ (right contact resistance) | *Schottky/ Tunnel* | ~20 | ~0.06 | ~54 / ~14 | ~17.88 / ~15.56 | ~30 / ~8 | ~1.13 / ~0.13 |
| $R_{total}$ | — | | ~0.3 | | ~33.11 / ~ 111.12 | | ~3.76 / ~1.67 |

## 4. Application: Contact Optimization by Thermal Annealing

To demonstrate the utility of the framework for contact optimization studies, we apply all three methods to sample D2, a bilayer $MoS_2$ device before and after thermal annealing as a widely used strategy for improving 2D semiconductor contacts.[5,8,9,12] Device D2 exhibits diode-like contact behavior before and after annealing, even though IV characterization (Figure 4a) shows a dramatic reduction in total resistance after annealing at a temperature of 200°C for 45 min in vacuum (1.4 x $10^{-3}$ mbar): from $R_{D2}(+0.2\ V) = 40.0\ M\Omega$ and $R_{D2}(-0.2\ V) = 285.7\ M\Omega$ before annealing to $R_{D2,ann}(+0.2\ V) = 2.5\ M\Omega$ and $R_{D2,ann}(-0.2\ V) = 3.5\ M\Omega$ after. Despite this improvement, the IV curves retain their diode-like shape, and the total resistance remains higher than that of the ohmic sample O1, indicating that contact barriers persist (see also additional data provided in the supplemental materials).

LBIC maps before annealing (Figure 4b) show photocurrent signals predominantly on the flake over both contacts, with a dominant negative signal at the upper-left and a smaller positive signal at the lower-right, attributable primarily to the PTE effect with minor photovoltaic and presumably tunnel contribution. After annealing (Figure 4c), the negative signal shifts further onto the left contact (bulk region of the flake), while the positive signal shifts into the gap and spreads across the full flake width, indicating a transition from PTE-dominated to photovoltaic-dominated response at the right contact. We explain the signal shift induced by the annealing process at the right contact by a reduction of the tunnel barrier. Prior to annealing, this tunnel barrier limits the effective contact area and prevents the formation of a well-defined Schottky junction across the full flake width. The reduction of the tunnel barrier resistance upon annealing enables an effective 2D semiconductor-metal junction to form across the entire flake width, resulting in an extended SCR.[16,45] This interpretation is consistent with the broader and more uniform positive photocurrent signal observed after annealing, which now spans nearly half of the length of the flake in the gap region. We attribute the shift of the negative photocurrent signal further onto the left contact to an improved contact between the bulk part of the $MoS_2$ flake and the left contact after annealing. This improvement enhances the thermal coupling at the metal-$MoS_2$ interface, strengthening the temperature gradient and thus the PTE effect. The signal is now predominantly located on the bulk region of the flake, where the higher optical absorption and larger Seebeck coefficient further promote the thermoelectric response. [4,5,37]

KPFM measurements after annealing (Figure 5) reveal a potential distribution qualitatively similar to that of sample D1, with a dominant, asymmetric potential drop at the contact edges that depends on voltage polarity and illumination. This confirms that contact barriers remain the dominant source of resistance even after annealing.

By comparing the LBIC and KPFM measurements for sample D2 after annealing, we can conclude that a dominant Schottky barrier is present at the right contact, evidenced by the photovoltaic photocurrent signal spanning the full flake length in the gap. The tunnel barrier at the right contact, which prior to annealing limited the effective contact area, is reduced by the annealing process, enabling the formation of a well-defined Schottky junction across the entire flake width resulting in an ohmic-like Schottky-contact. At the left contact, the PTE effect still dominates, consistent with an overall improved ohmic-like contact behavior of the flake after annealing.

## 5. Conclusion

We have presented an integrated characterization framework combining IV measurements, LBIC mapping, and bias-dependent KPFM for the comprehensive analysis of metal contacts to few-layer $MoS_2$ devices under ambient conditions. Each method contributes essential and complementary information: IV measurements quantify macroscopic transport and total resistance; LBIC maps the spatial distribution and mechanistic origin of internal electric fields; and KPFM under applied bias directly visualizes the relative magnitudes and locations of contact barriers with nanoscale resolution.

Applied to devices with ohmic-like and diode-like contact behavior, the framework unambiguously identifies whether the dominant barrier is of Schottky or tunnel type and resolves the asymmetry between the two contacts in a given device. Applied to a thermally annealed device, it reveals that annealing significantly reduces total resistance and differentially modifies the two contact barriers, yet contact barriers remain the dominant source of resistance even after optimization.

The methodology is straightforward to implement, requires no cryogenic or vacuum environment, and does not require a gate electrode. It is directly transferable to other 2D semiconductor–metal systems and to a broader range of contact optimization strategies, including surface doping, metal selection, and strain engineering. Beyond contact characterization, the framework could be extended to investigate heterojunctions, grain boundaries, and other laterally inhomogeneous structures in 2D materials. We anticipate that the combined IV/LBIC/KPFM approach will serve as a practical and broadly applicable diagnostic tool for the rational engineering of low-resistance contacts in 2D semiconductor devices.

**Figure 1**

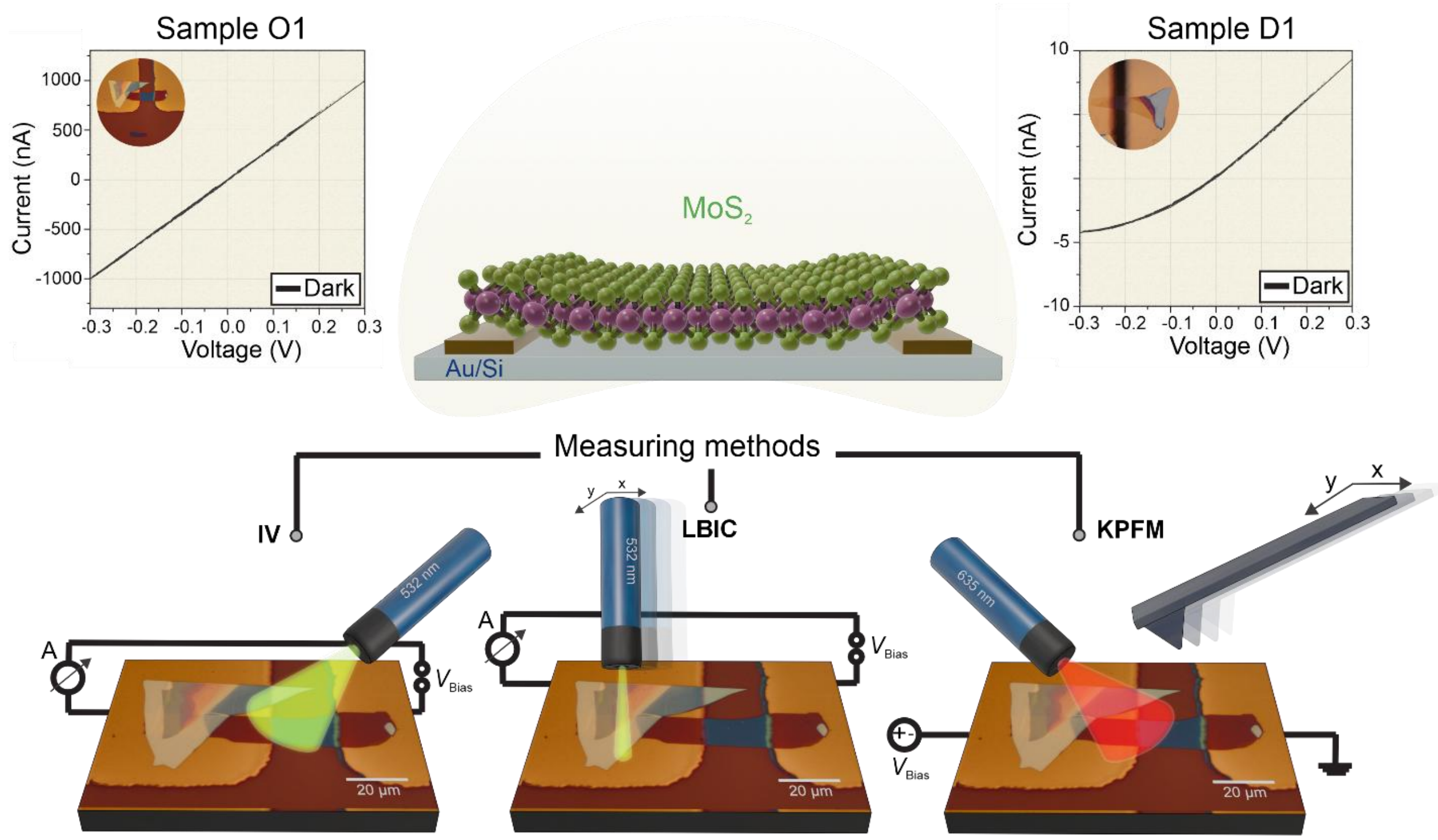

**Figure 1.** $MoS_2$ devices and complementary measurement methods. Center: Schematic side view of the $MoS_2$ flake on top of Cr/Au contacts on a Si/$SiO_2$ or glass substrate. Upper left and right: IV characteristics in the dark of sample O1 (ohmic-like) and sample D1 (diode-like). Insets: optical micrographs of the $MoS_2$ flakes. Lower line: Schematic overview of the three complementary measurement methods. IV:; current is measured as a function of applied bias; for illuminated IV characterization a defocused laser beam globally excites the $MoS_2$ flake LBIC: a focused laser beam scans the sample in ~2 µm steps; photocurrent and reflected laser intensity are recorded simultaneously at each position; a bias voltage can be applied. KPFM: the CPD is measured by the tip while scanning the sample in ~100 nm steps; a static bias is applied and the $MoS_2$ flake can optionally be globally illuminated.

## Figure 2

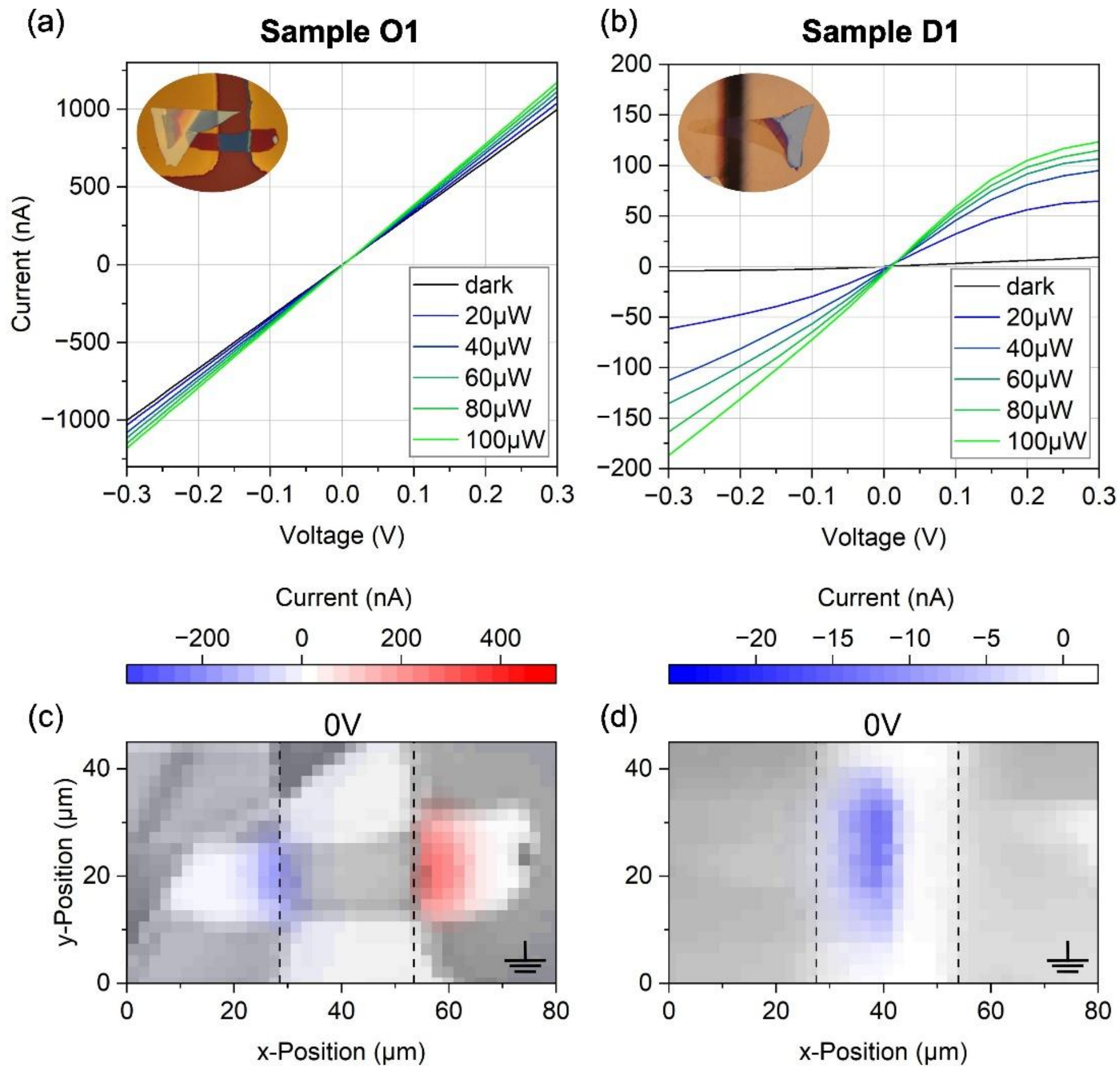

**Figure 2.** IV and LBIC characterization of samples O1 and D1. IV characteristics of (a) sample O1 and (b) sample D1 in the dark and under illumination at laser powers of 20, 40, 60, 80, and 100 μW. Insets: optical micrographs of the $MoS_2$ flakes. LBIC maps at 0 V bias of (c) sample O1 and (d) sample D1. Photocurrent (linear color scale) is overlaid on the reflected-light image (grayscale). Dashed black lines mark the contact edges. [Excitation: $E_{Laser}$ = 2.33 eV (λ = 532 nm), P = 100 μW]

**Figure 3**

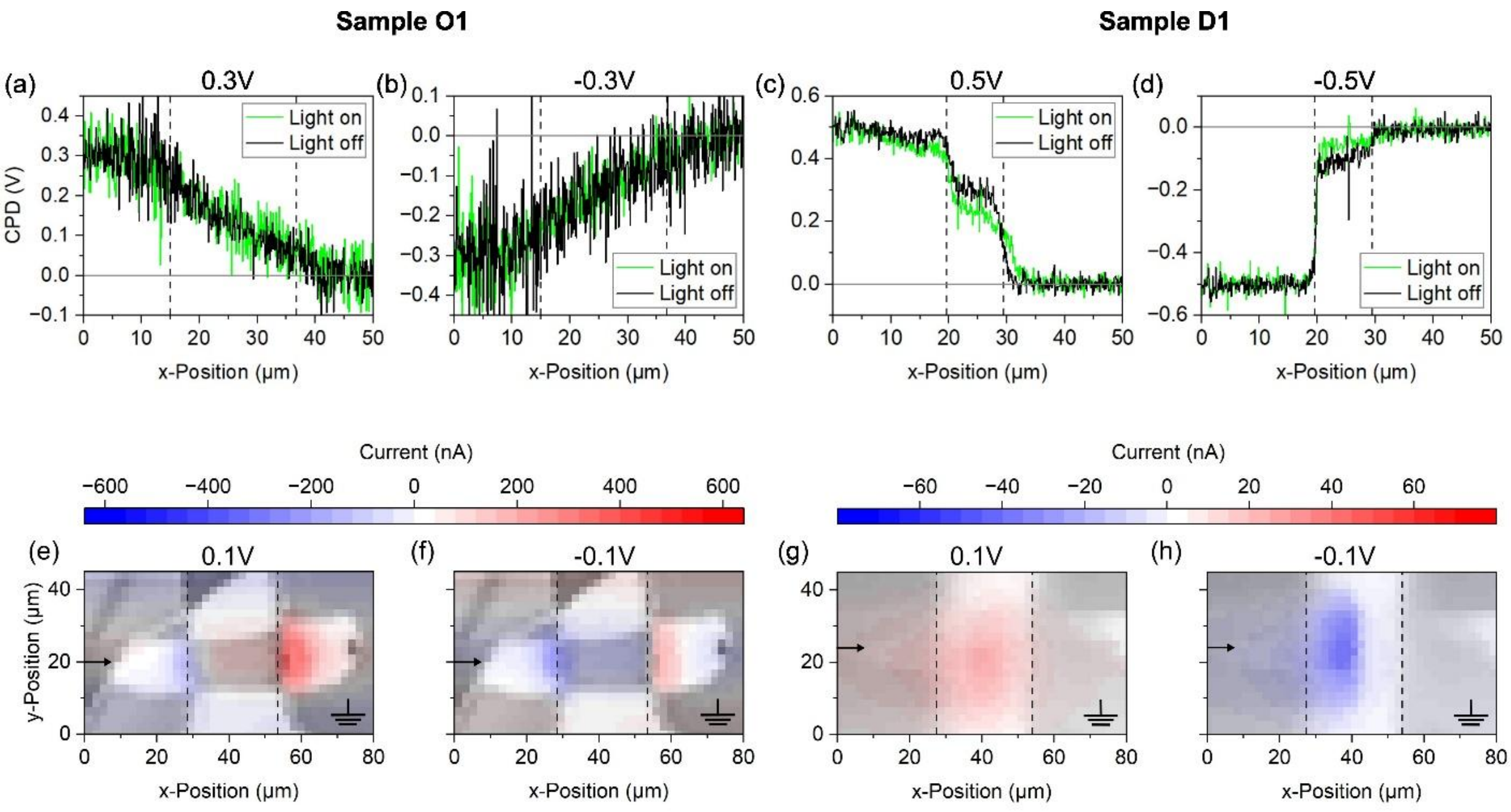

**Figure 3.** Bias-dependent KPFM surface voltage profile lines and LBIC maps for samples O1 and D1. Bias dependent surface voltage profile lines (CPD corrected by 0 V reference CPD scan) measured at +0.3 V and −0.3 V for sample O1 (a, b) and at +0.5 V and −0.5 V for sample D1 (c, d), in the dark and under illumination. The profile line position is representative of all y-positions on the $MoS_2$ flake. LBIC maps of (e, f) sample O1 and (g, h) sample D1 at bias voltages of +0.1 V and −0.1 V. Photocurrent (linear color scale) is overlaid on the reflected-light image (grayscale); maps are corrected for the dark current at the corresponding bias. The right contact is set to ground potential. [KPFM: $E_{Laser}$ = 1.95 eV (λ = 635 nm), P = 5 mW; LBIC: $E_{Laser}$ = 2.33 eV (λ = 532 nm), P = 100 μW]

**Figure 4**

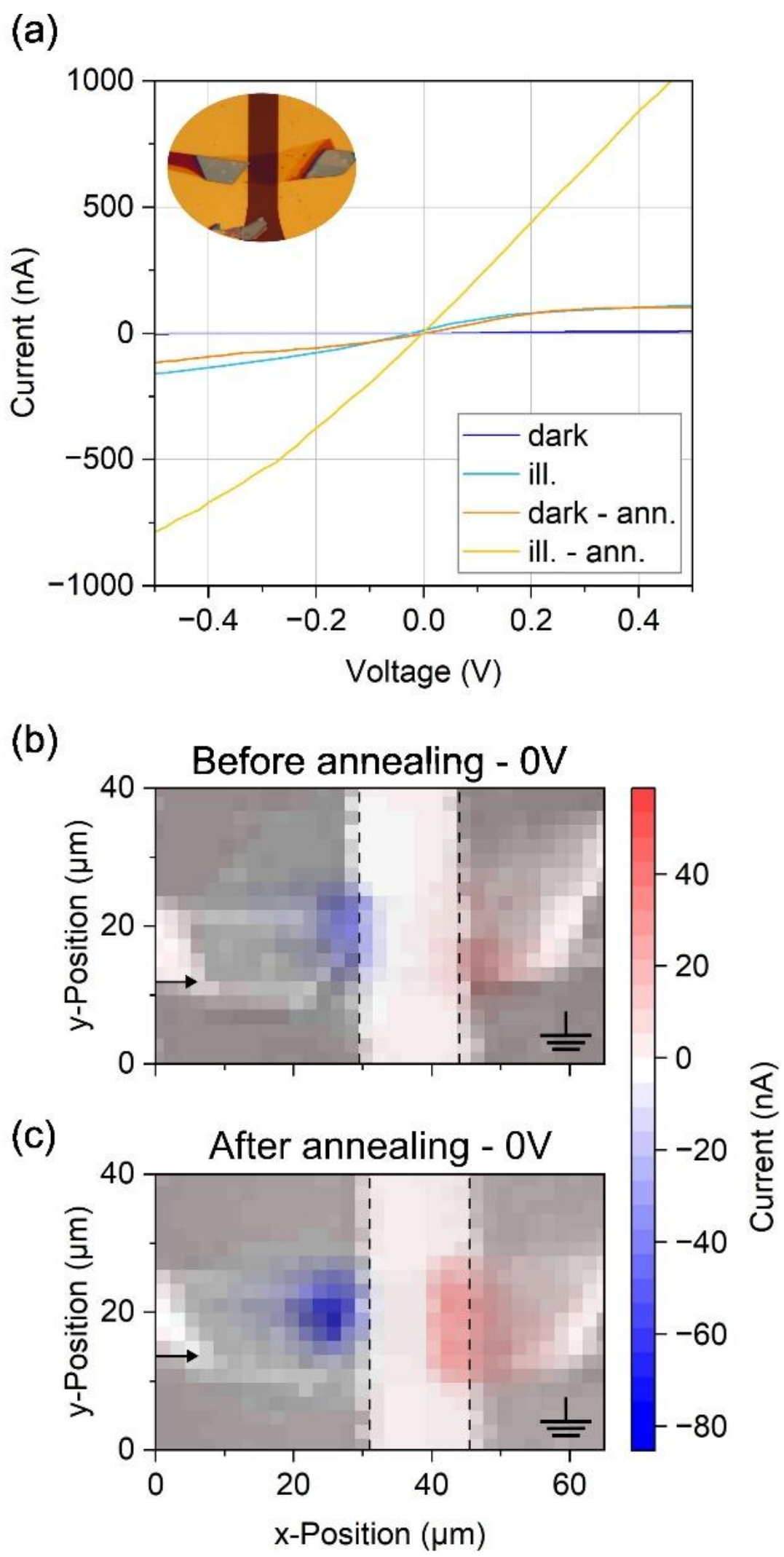

**Figure 4.** Effect of thermal annealing on IV and LBIC characteristics of sample D2. (a) IV characteristics before (blue) and after (orange) thermal annealing, in the dark and under illumination. Inset: optical microscopy image of the $MoS_2$ flake. LBIC maps at 0 V bias (b) before and (c) after thermal annealing. Photocurrent (linear color scale) is overlaid on the reflected-light image (grayscale). Dashed vertical lines mark the contact edges. Black arrows indicate the positions at which the bias induced surface voltage profile lines in Figure 5 are extracted. [Vacuum annealing conditions: $T_{ann}$ = 200°C, $t_{ann}$ = 45 min, $p_{ann}$ = 1.4 x $10^{-3}$ mbar; Excitation: $E_{Laser}$ = 2.33 eV (λ = 532 nm), P = 100 µW]

**Figure 5**

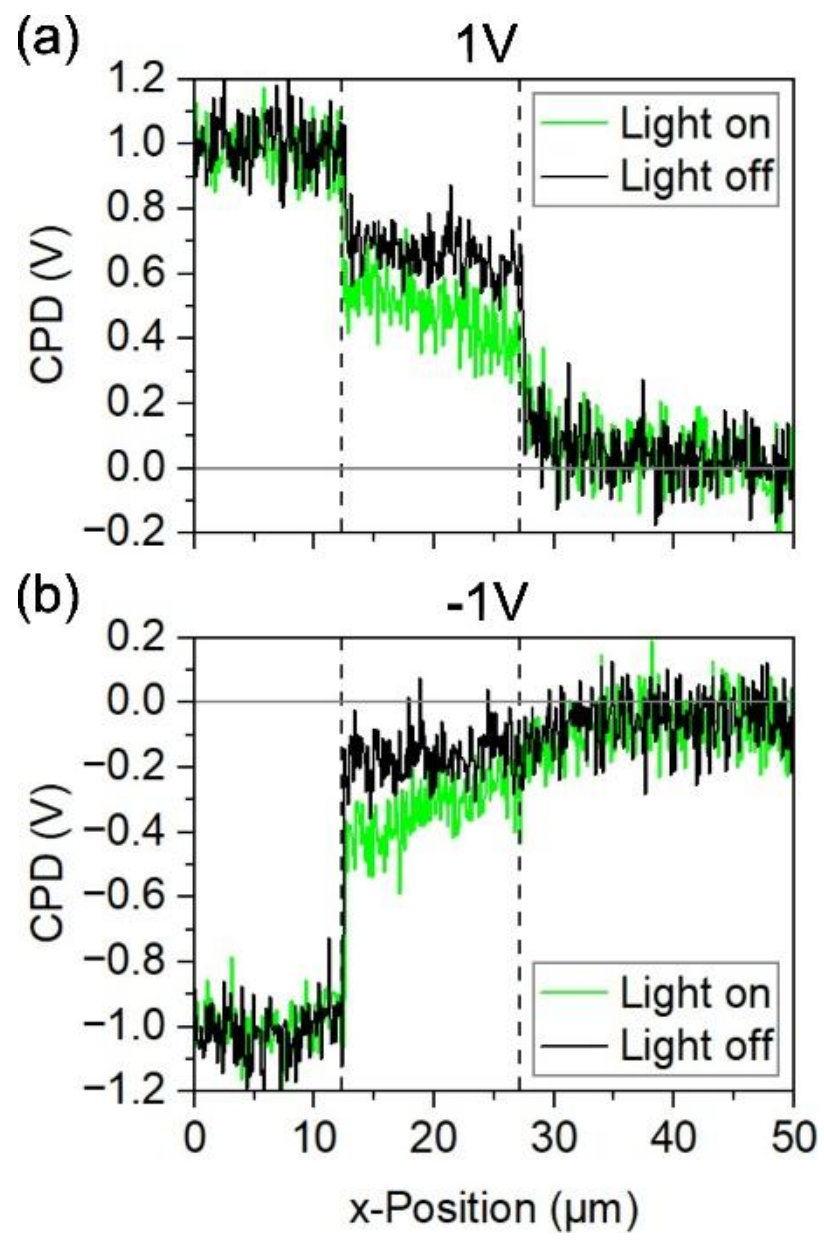

**Figure 5.** Bias-dependent KPFM surface voltage profile lines of thermally annealed sample D2. Bias induced surface voltage profile lines measured at positive and negative bias voltages, in the dark and under illumination, after thermal annealing. Profile line positions are marked by arrows in Figure 4c. [KPFM: $E_{Laser}$ = 1.95 eV (λ = 635 nm), P = 1 mW]

## Acknowledgements

This publication is part of the project Piezo-photomotion: light-driven nano piezo-propulsion and agitation with project number VI.Vidi.193.020 of the research programs Vidi and Aspasia, which is financed by the Dutch Research Council (NWO). This project has received funding from the European Union's Horizon Europe research and innovation program under grant agreement No. 101130224 "JOSEPHINE". The authors acknowledge funding from the University of Twente – University of Münster collaboration fund.

## Data Availability Statement

The data that support the findings of this study are available from the corresponding author upon reasonable request.

## Declaration of Generative AI and AI-Assisted Technologies

During the preparation of this work, the authors used large language models to refine the manuscript regarding grammar, readability, and clarity. After using these tools, the authors reviewed and edited the content as needed and take full responsibility for the content of the publication. No additional use of artificial intelligence was made during data acquisition, analysis, interpretation, or other aspects of this study.

# Supplementary Material for the article:

# Identifying Contact Barrier Types in Few-Layer $MoS_2$ Devices Using Correlative IV, LBIC, and Bias-Dependent KPFM

Ariane Ufer[1,2], Zeinab Eftekhari[3], Benjamin Mayer[1], Hendrik Lambers[1,2], Hubert J. Krenner[1], Rebecca Saive[3,*], Ursula Wurstbauer[1,2,*]

[1] Institute of Physics, University of Münster, Münster, Germany

[2] Center for Soft Nanoscience (SoN), University of Münster, Münster, Germany

[3] MESA+ Institute for Nanotechnology, University of Twente, Enschede, The Netherlands

*Corresponding authors: wurstbauer@uni-muenster.de; r.saive@utwente.nl

## Content

# 1. Overview of samples

## 1.1 Optical micrographs

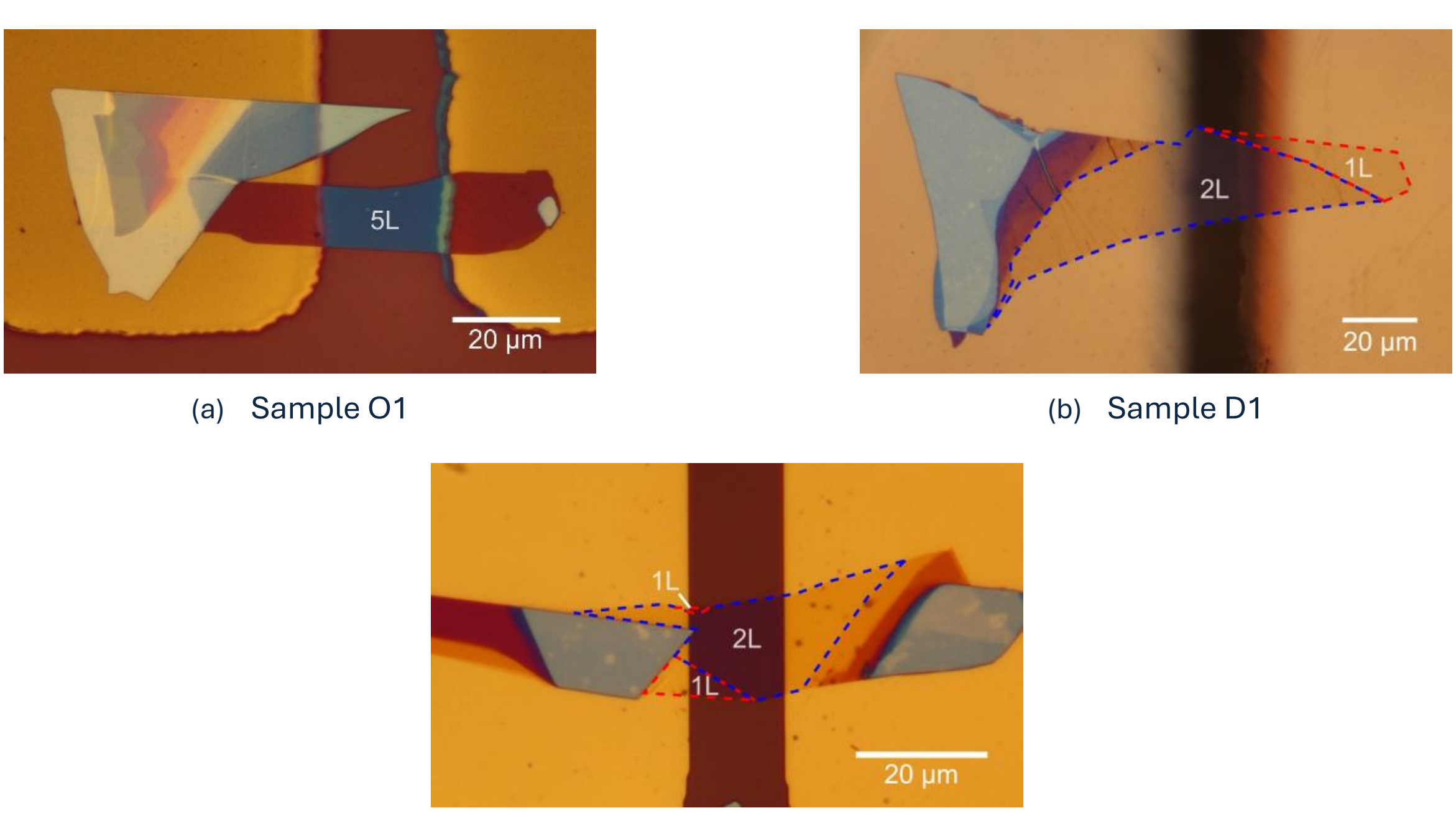

(a) Sample O1 (b) Sample D1

(c) Sample D2

*Figure 1* Optical microscope image of the samples O1 in (a), D1 in (b) and D2 in (c). In each image the gold contacts are shown with the gap in between. On top of the contacts and the gap the individual $MoS_2$ flakes of the samples are located. The $MoS_2$ flakes were fabricated by micromechanical exfoliation and transferred to the contacts using the dry viscoelastic stamping technique. The different layer thicknesses of the thin flake parts are indicated as 1L for a monolayer, 2L for a bilayer and 5L for five layers. These layer thicknesses are determined by Ramans spectroscopy. The substrate of the samples O1 and D2 is $Si/SiO_2$ and the substrate of the sample D1 is glass. The gold contacts of the samples O1 and D1 were fabricated using a shadow mask and the gold contacts of the sample D2 using UV photolithography.

# 2. Overview results IV measurements (dark and illuminated)

## 2.1 IV curves of Sample O1

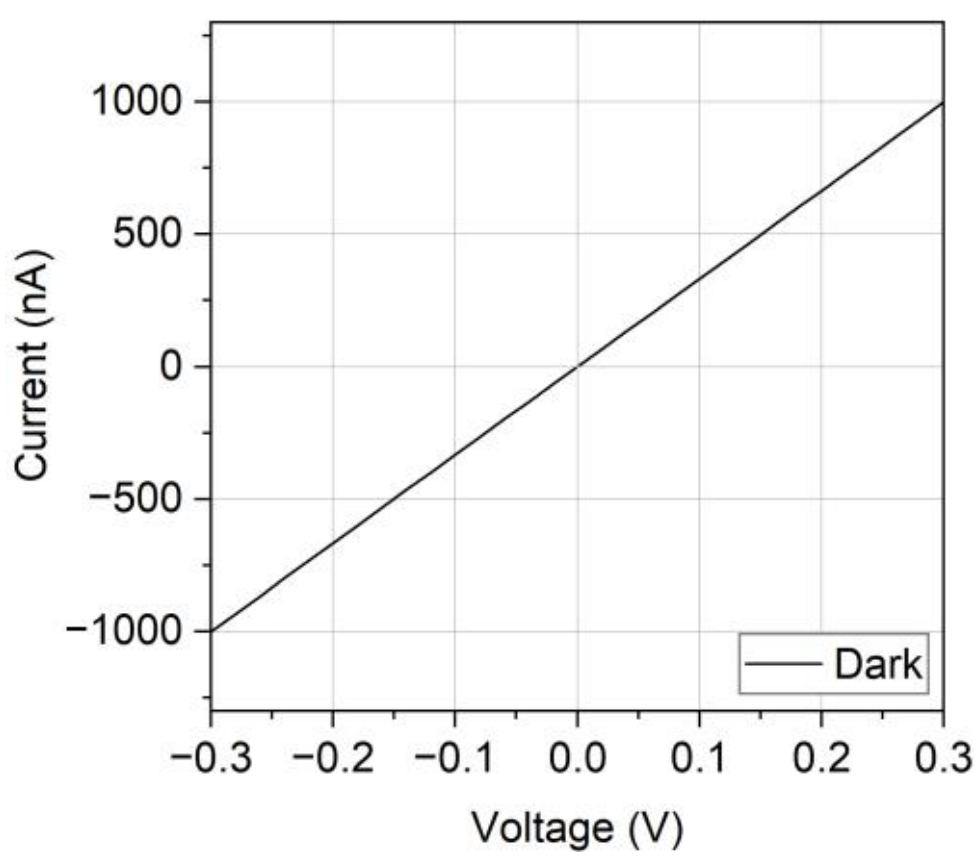

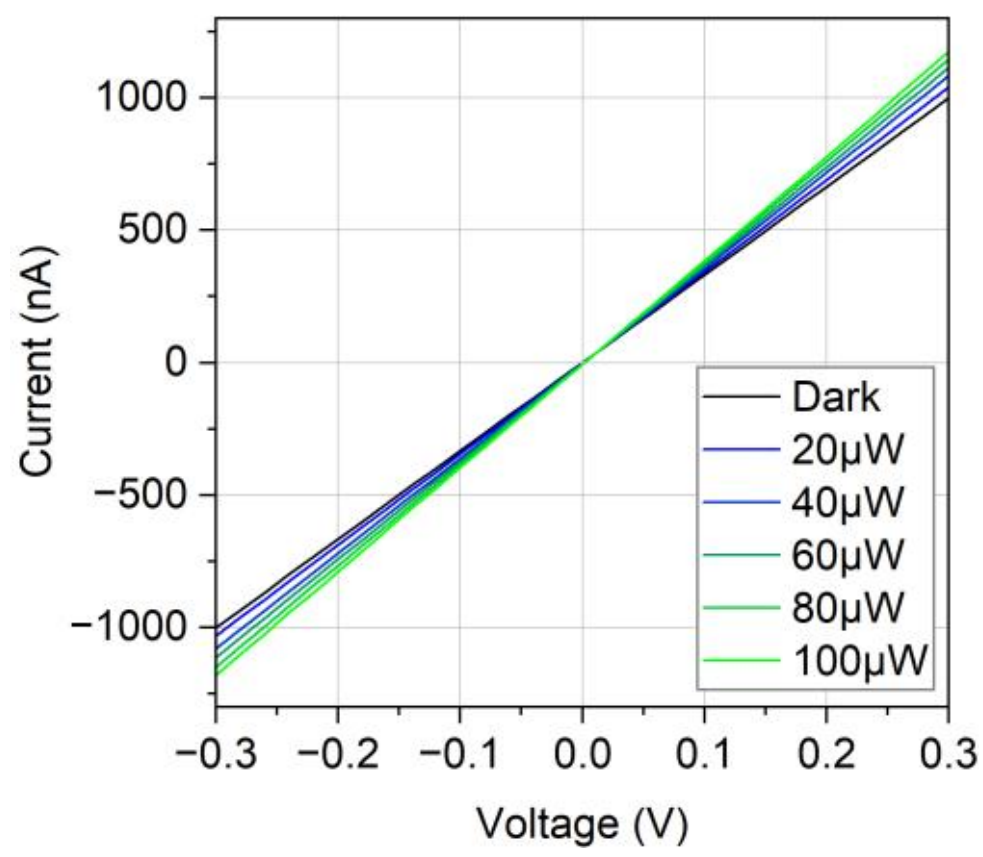

*Figure 2* IV characteristics of sample O1 (left) without light exposure and (right) without light exposure in comparison with the measurements with light exposure at different laser powers of 20, 40, 60, 80 and 100 μW. The sample is globally excited by laser light with a wavelength of 532 nm and thus with an energy of $E_{laser}$ = 2.33 eV that is well above the band gap of $MoS_2$. The current values are presented as a function of voltages in the range of −0.3V to 0.3 V. A linear and symmetric function with large current values compared to the other samples is observed for all different light exposure settings. The measurements are carried out at ambient conditions. Sample O1 presents an ohmic behavior. Therefore, it can be assumed that the junctions from the $MoS_2$ flake with each of the two gold contacts are ohmic contacts with low resistances.

## 2.2 IV curves of Sample D1

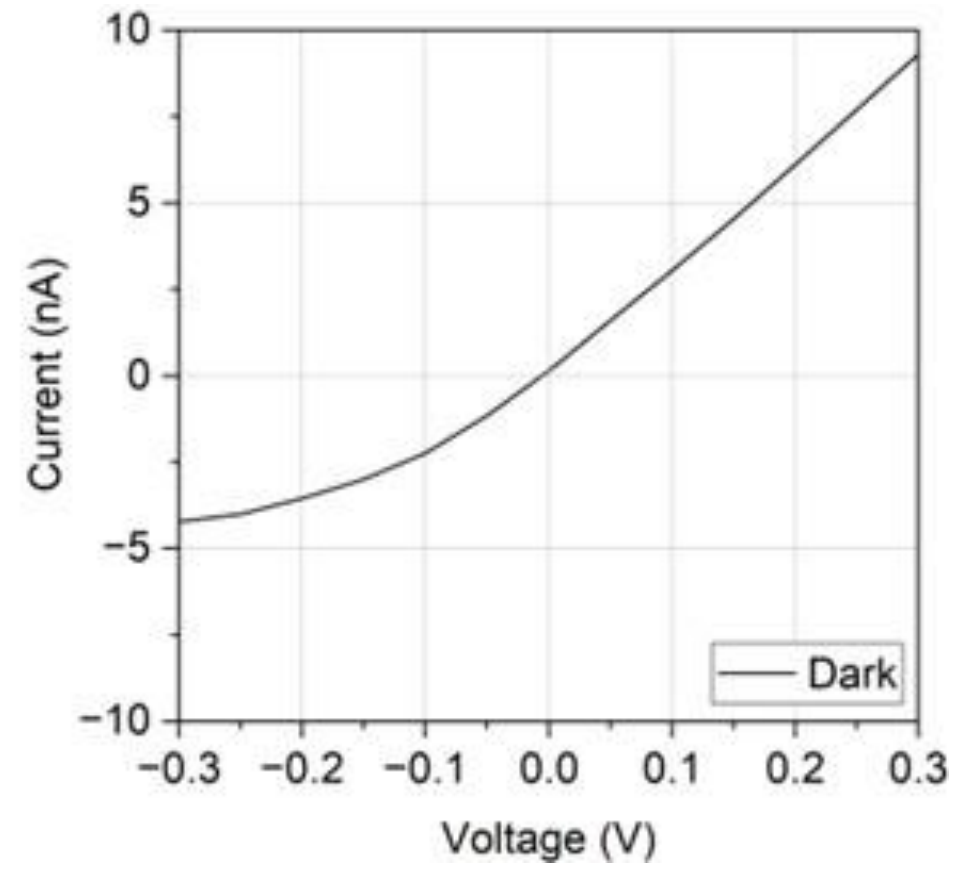

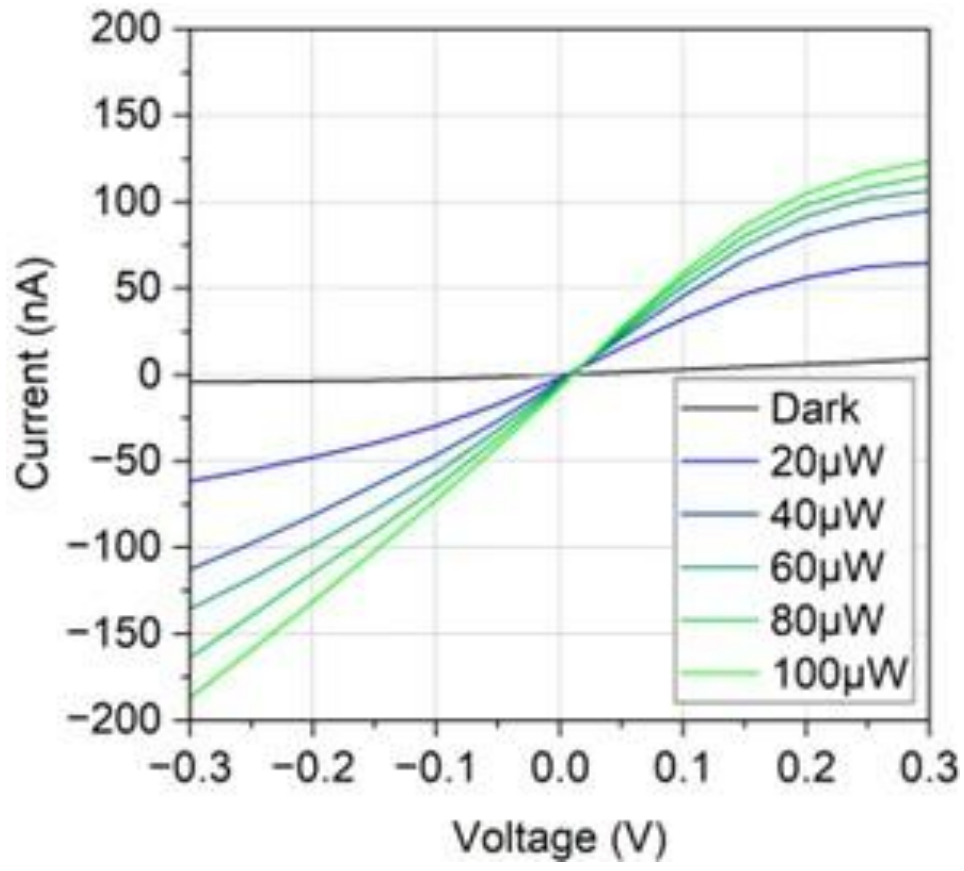

*Figure 3* IV characteristics of sample D1 (left) without light exposure and (right) without light exposure in comparison with the measurements with light exposure at different laser powers of 20, 40, 60, 80 and 100 μW. The sample is globally excited by laser light with a wavelength of 532 nm and thus with an energy of $E_{laser}$ = 2.33 eV that is well above the band gap of $MoS_2$. The current values are presented as a function of voltages in the range of −0.3V to 0.3 V. The IV characteristics for the different light exposure settings show a non-linear and asymmetric behavior for positive and negative voltages with saturations for larger voltages. The current values significantly depend on the light exposure setting but are generally low compared to sample O1. The measurements are carried out at ambient conditions. Sample D1 presents a non-ohmic behavior with voltage-dependent resistances. Therefore, it can be assumed that the junctions from the $MoS_2$ flake with each of the two gold contacts exhibit asymmetric contact barriers resulting in large resistances.

## 2.3 IV curves of sample D2 before annealing

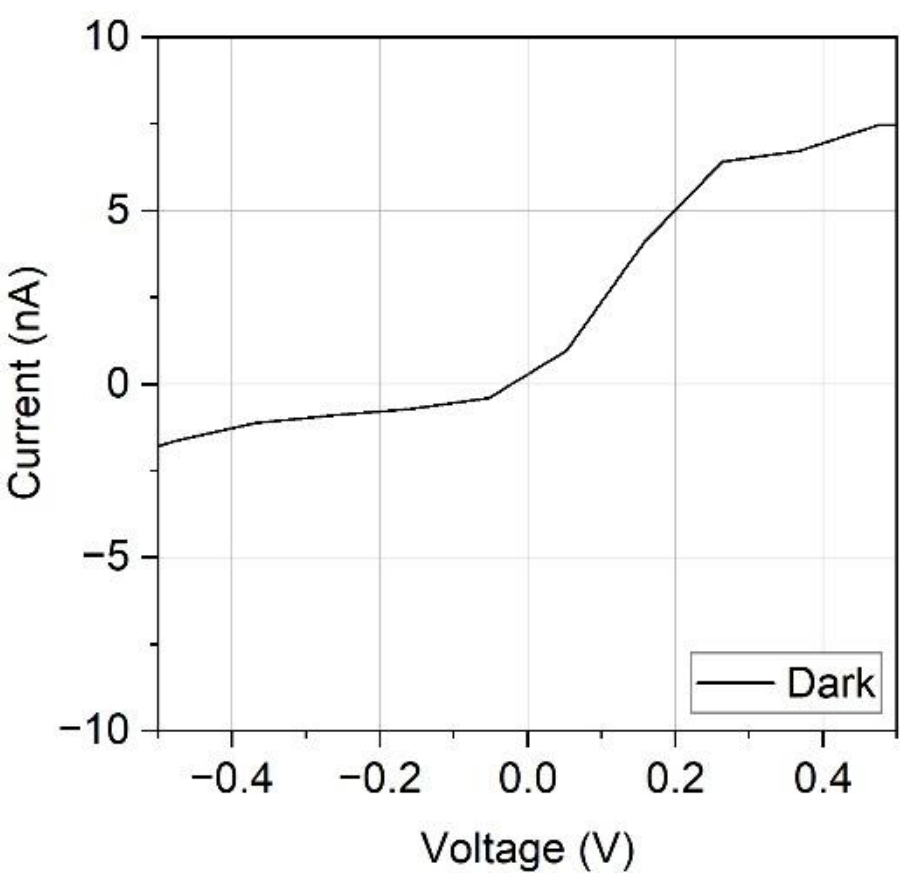

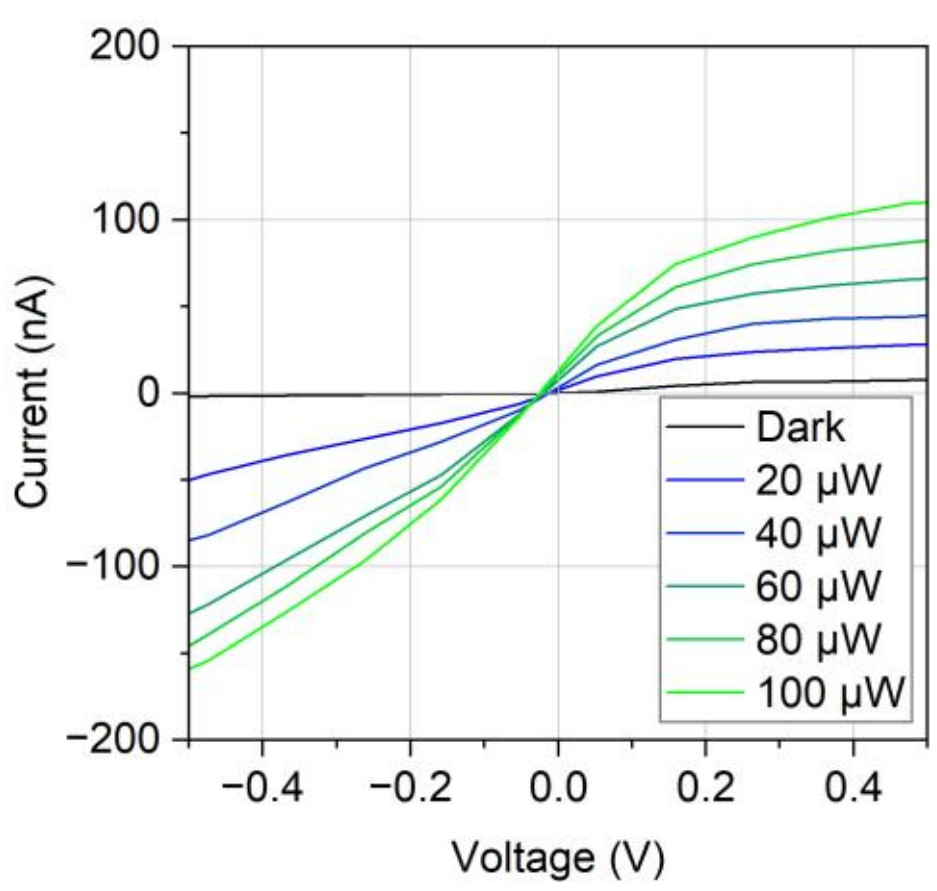

*Figure 4* IV characteristics of sample D2 before annealing (left) without light exposure and (right) without light exposure in comparison with the measurements with light exposure at different laser powers of 20, 40, 60, 80 and 100 μW. The sample is globally excited by laser light with a wavelength of 532 nm and thus with an energy of $E_{laser}$ = 2.33 eV that is well above the band gap of $MoS_2$. The current values are presented as a function of voltages in the range of −0.5V to 0.5 V. The measurements are carried out at ambient conditions. Asymmetric IV curves with saturation and small current values compared to the measurement after annealing are observed. This leads to the conclusion of asymmetric contact barriers similar to the IV results of sample D1.

## 2.4 IV curves of sample D2 after annealing

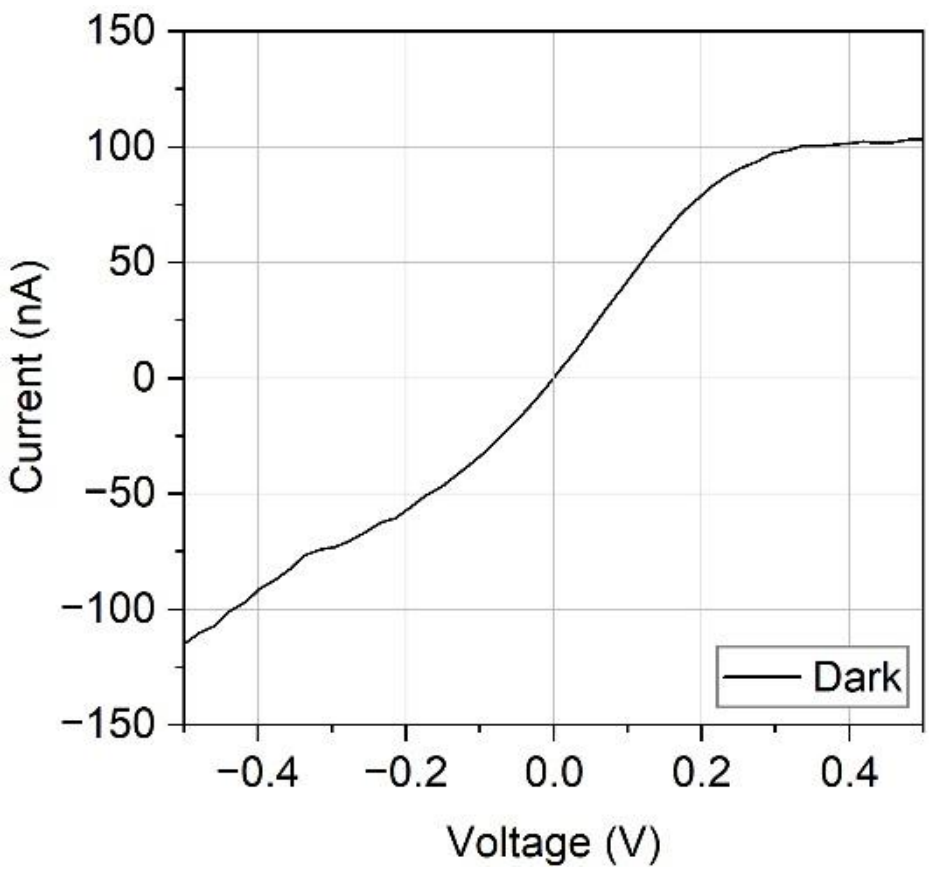

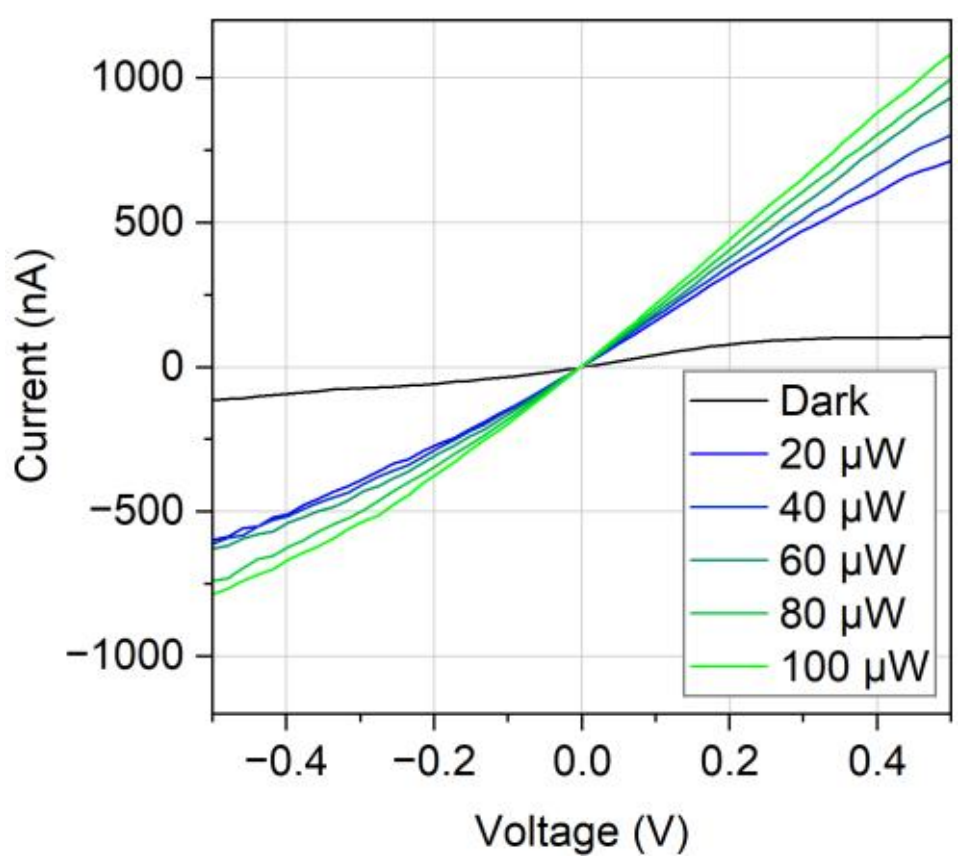

*Figure 5* IV characteristics of sample D2 after annealing (left) without light exposure and (right) without light exposure in comparison with the measurements with light exposure at different laser powers of 20, 40, 60, 80 and 100 µW. The sample is globally excited by laser light with a wavelength of 532 nm and thus with an energy of $E_{laser}$ = 2.33 eV that is well above the band gap of $MoS_2$. The current values are presented as a function of voltages in the range of −0.5V to 0.5 V. The measurements are carried out at ambient conditions. After annealing the IV characteristics are more symmetric with reduced non-linearity and larger currents values compared to the measurement before annealing. This indicates an improved contacts with reduced contact resistances.

## 2.5 Summary of the results of the measured IV characteristics of Sample O1 and D1:

| **Sample O1** | **Sample D1** |
|---|---|
| Linear IV characteristics | Asymmetric IV curves with saturation (Diode-like) |
| Large currents → low contact resistances | Small currents → high contact resistances |
| Symmetric and ohmic contact behavior | Asymmetric contacts with contact barriers |

## 2.6 Summary of the results of the IV measurements of Sample D2:

| Before annealing | After annealing |
|---|---|
| Asymmetric IV curves with saturation | More symmetric IV characteristics with reduced non-linearity |
| Small currents → high contact resistances | Large currents → lower contact resistances |
| Asymmetric contact barriers | Indicating improved contact transparency |
| Similar results as for IV curves of sample D1 | |

# 3. Localized laser beam induced current (LBIC) measurements

## 3.1 Photocurrent data of sample O1 and schemata of according bands

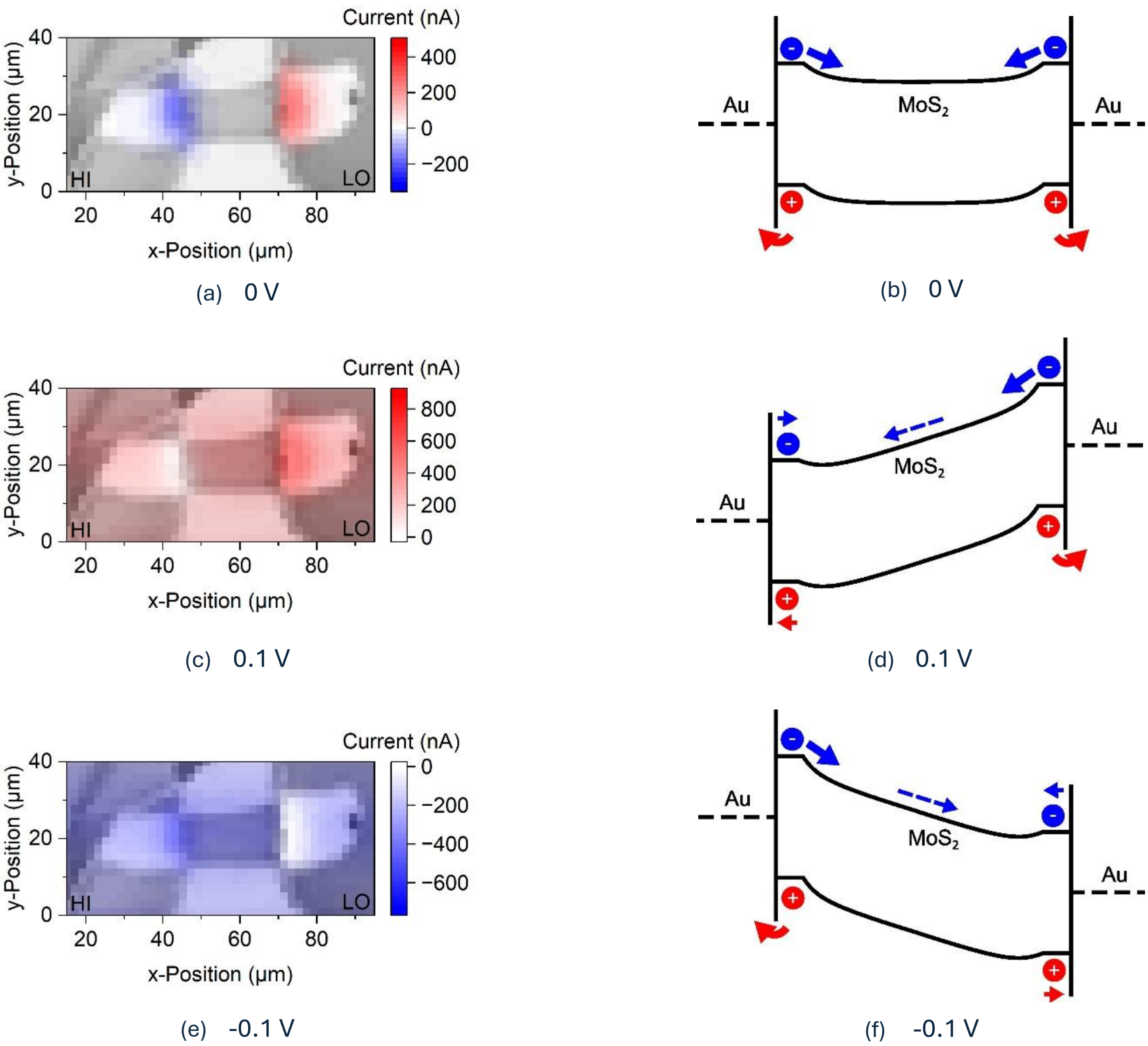

*Figure 6* Photocurrent results of sample O1. The laser light with a wavelength of 532 nm is focused on the sample and scanned over it in steps of 2 µm. The photocurrent maps are shown on the left side and corresponding band models on the right side at voltages of 0V in (a) and (b), 0.1V in (c) and (d) and −0.1V in (e) and (f). A blue signal in the heat maps corresponds to a negative current and a red signal to a positive current. The x- and y-axis shows the lateral spatial dimensions of the sample. In transparent grey colors the reflectance image of the contacts and the flake are superimposed. HI and LO indicate the connection of the contacts to the outputs (high and low) of the source meter unit. The schemata show the charge carrier movement that is induced by internal electric fields and local light exposure in the VB and CB. The size of the arrows corresponds to the size of the measured currents and thus to the size of the electrical field. The dashed arrows demonstrate the background current at an applied voltage. Sample O1 shows an approximately symmetric photocurrent signal, which is mainly located on both contacts. It is concluded that the photocurrent is dominated by the photothermoelctric effect. The photocurrent values are large compared to sample D1. The observed behavior is explained by approximately symmetric contacts with no Schottky barriers.

## 3.2 Photocurrent data of sample D1

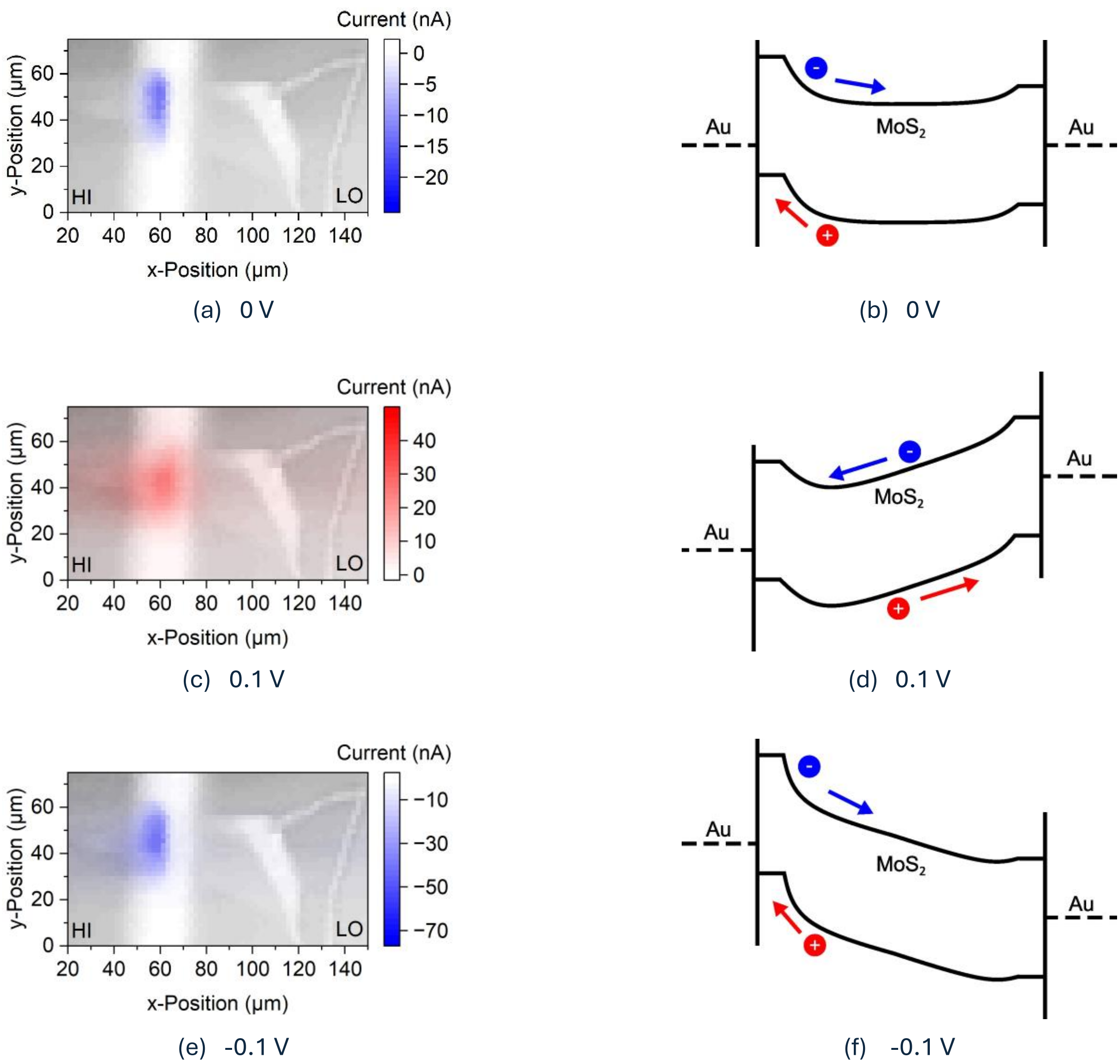

*Figure 7* Photocurrent results of sample A1. The laser light with a wavelength of 532 nm is focused on the sample and scanned over it in steps of 2 µm. The photocurrent maps are shown on the left side and corresponding band models on the right side at voltages of 0V in (a) and (b), 0.1V in (c) and (d) and −0.1V in (e) and (f). A blue signal in the heat maps corresponds to a negative current and a red signal to a positive signal. The x- and y-axis shows the lateral spatial dimensions of the sample. In transparent grey colors the reflectance image of the contacts and the flake are superimposed. HI and LO indicate the connection of the contacts to the outputs (high and low) of the source meter unit. The schemata show the charge carrier movement that is induced by internal electric fields and local light exposure in the VB and CB. The size of the arrows corresponds to the size of the measured currents and thus to the size of the electrical field. The bending of the bands also depicts the internal electric fields. Sample D1 exhibits asymmetric photocurrent signals within the gap implying asymmetric contact barriers at the contacts. The photocurrent signal is dominated by the photovoltaic effect. The photocurrent values are small compared to sample O1. For sample D1 a Schottky barrier at the left contact and a potential tunnel barrier at the right contact can be concluded.

### 3.3 Summary of the results from LBIC measurements of sample O1 and D1:

| Sample O1 | Sample D1 |
|---|---|
| Photocurrent signal is mainly located on both contacts | Photocurrent signal is located in the gap between the contacts |
| Dominated by the PTE effect at the gold-$MoS_2$ junctions | Dominated by the photovoltaic effect at the left Schottky barrier |
| Almost symmetric signal at both contacts | Asymmetric signal that dominated at the left contact |
| Large photocurrent values | Small photocurrent values |
| Indicating small barriers and almost symmetric ohmic contacts | Indicating large and asymmetric contact barriers, a Schottky barrier at the left and possibly a tunnel barrier at the right contact |

## 3.4 Local laser beam induced current (LBIC) measurements before and after annealing of sample D2

**Photocurrent data sample D2 before annealing**

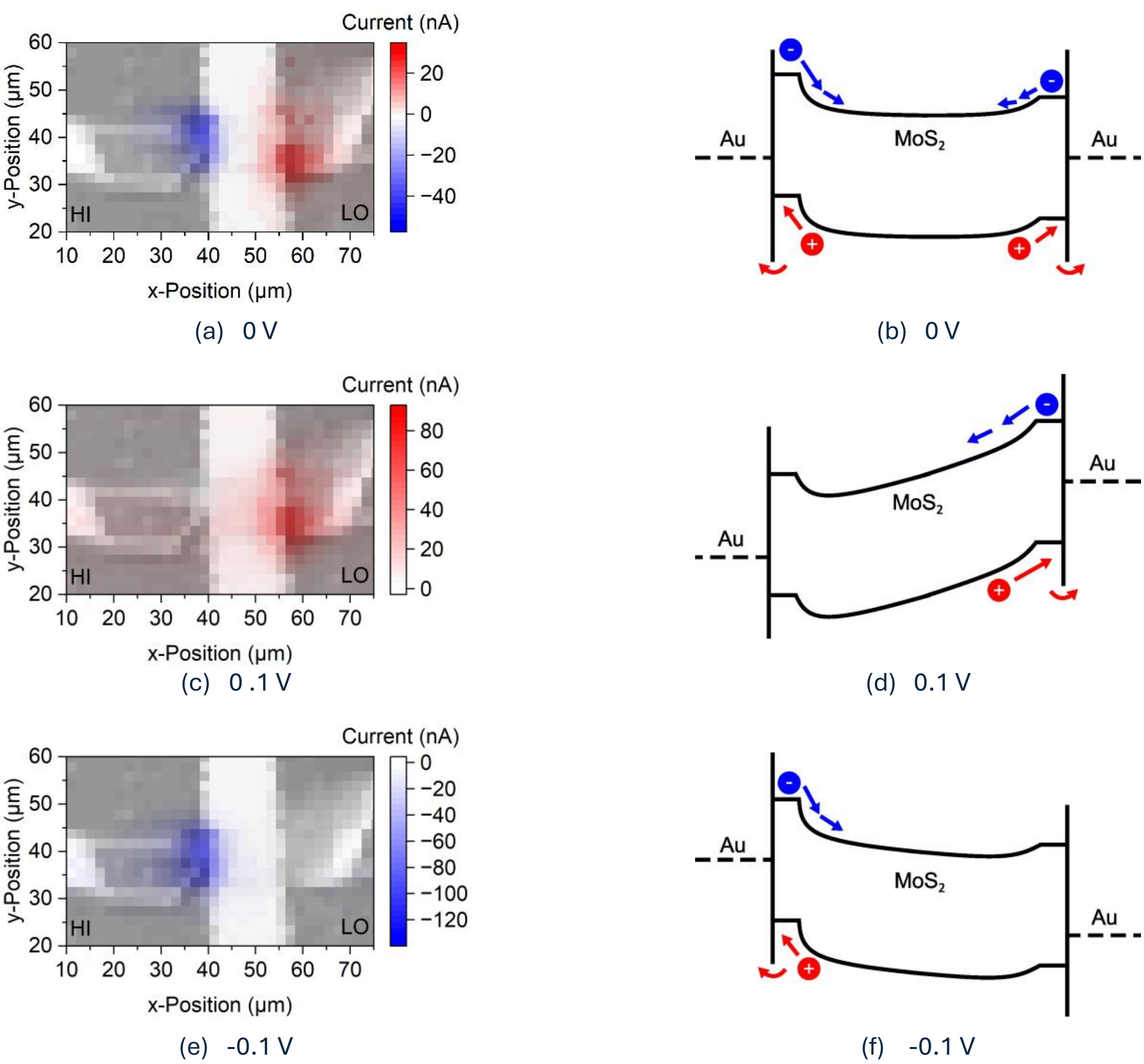

*Figure 8* Photocurrent results of sample D2 before annealing. The laser light with a wavelength of 532 nm is focused on the sample and scanned over it in steps of 2 µm. The photocurrent maps are shown on the left side and corresponding band models on the right side at voltages of 0V in (a) and (b), 0.1V in (c) and (d) and −0.1V in (e) and (f). A blue signal in the heat maps corresponds to a negative current and a red signal to a positive signal. The x- and y-axis shows the lateral spatial dimensions of the sample. In transparent grey colors the reflectance image of the contacts and the flake are superimposed. HI and LO indicate the connection of the contacts to the outputs (high and low) of the source meter unit. The schemata show the charge carrier movement that is induced by internal electric fields and local light exposure in the VB and CB. The size of the arrows corresponds to the size of the measured currents and thus to the size of the electrical field. The bending of the bands also depicts the internal electric fields. To present the contribution of the PTE effect on the contacts and the photovoltaic effect in the gap close to the contacts in each case two arrows are sketched. Sample D2 before annealing exhibits asymmetric photocurrent signals especially regarding the current values. The photocurrent is mainly located on the part of the flake on top of the contacts indicating a main contribution by the PTE effect. A smaller contribution by the photovoltaic effect can be seen on the part of the flake within the gap close to the contacts indicating Schottky barriers and the photovoltaic effect. The photocurrent values are small compared to sample O1 but larger than those of sample D1. The photocurrent results are very different than those for sample D1.

**Photocurrent data sample D2 after annealing**

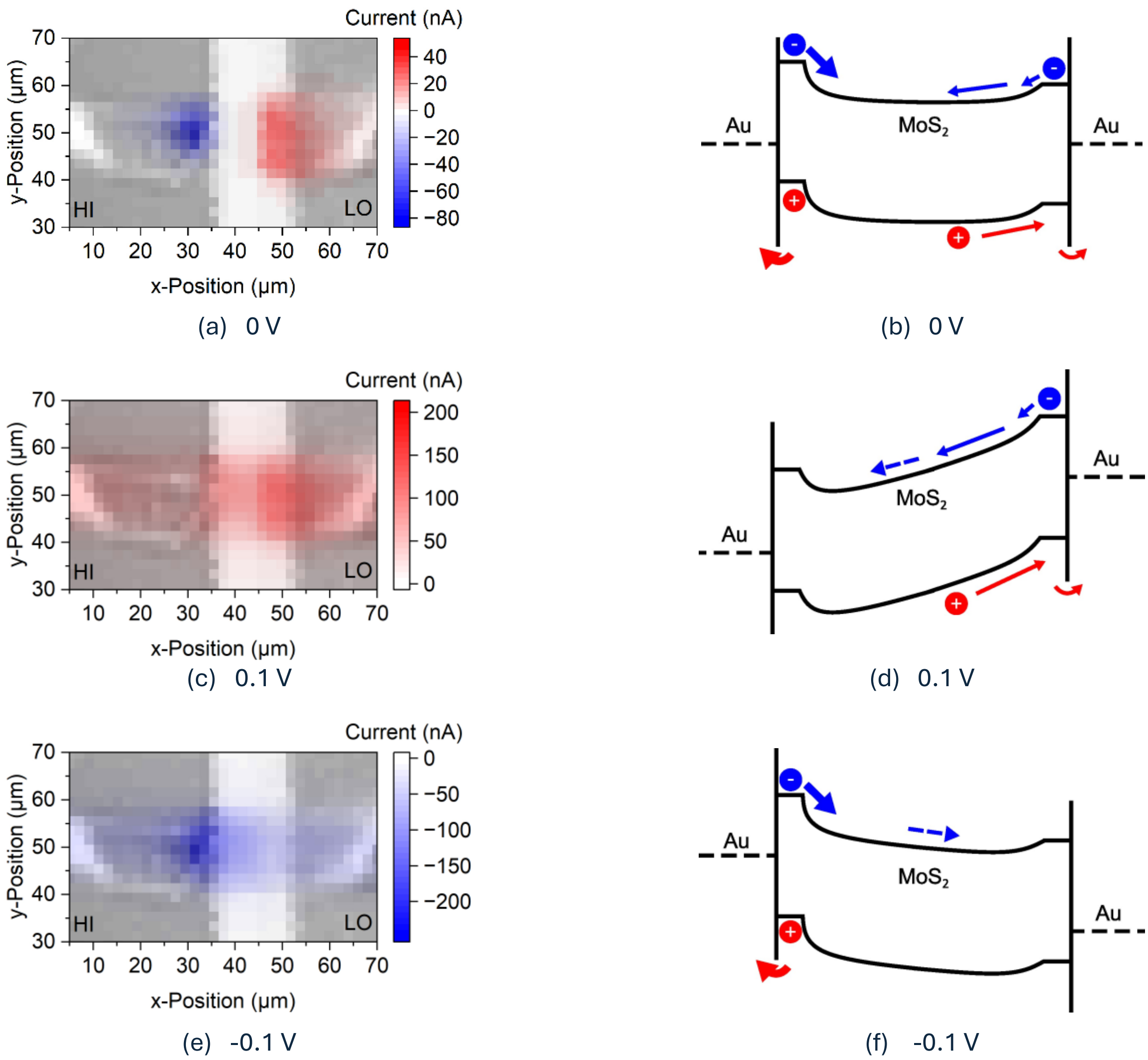

*Figure 9* Photocurrent results of sample D2 after annealing. The photocurrent maps are shown on the left side and corresponding band models on the right side at voltages of 0V in (a) and (b), 0.1V in (c) and (d) and −0.1V in (e) and (f). A blue signal in the heat maps corresponds to a negative current and a red signal to a positive signal. The x- and y-axis shows the lateral spatial dimensions of the sample. In transparent grey colors the reflectance image of the contacts and the flake are superimposed. HI and LO indicate the connection of the contacts to the outputs (high and low) of the source meter unit. The schemata show the charge carrier movement that is induced by internal electric fields and local light exposure in the VB and CB. The size of the arrows corresponds to the size of the measured currents and thus to the size of the electrical field. The bending of the bands also depicts the internal electric fields. To present the contribution of the PTE effect on the contacts and the photovoltaic effect in the gap close to the contacts in each case two arrows are sketched. Sample D2 after annealing exhibits asymmetric photocurrent signals regarding the current values and positions. Compared to the results before annealing the negative photocurrent signal shifted a bit further on to the left contact, whereas the observed positive photocurrent signal shifted a bit further into the gap. This indicates a dominated PTE effect at the left gold-$MoS_2$ junctions and a dominating photovoltaic effect close to the right contact. After annealing the measured photocurrent values are increased, which implies improved contact transparency. The measured photocurrent values are still significantly smaller than those of sample O1.

# 4. KPFM measurements with applied voltage

## 4.1 KPFM potential data of Sample O1

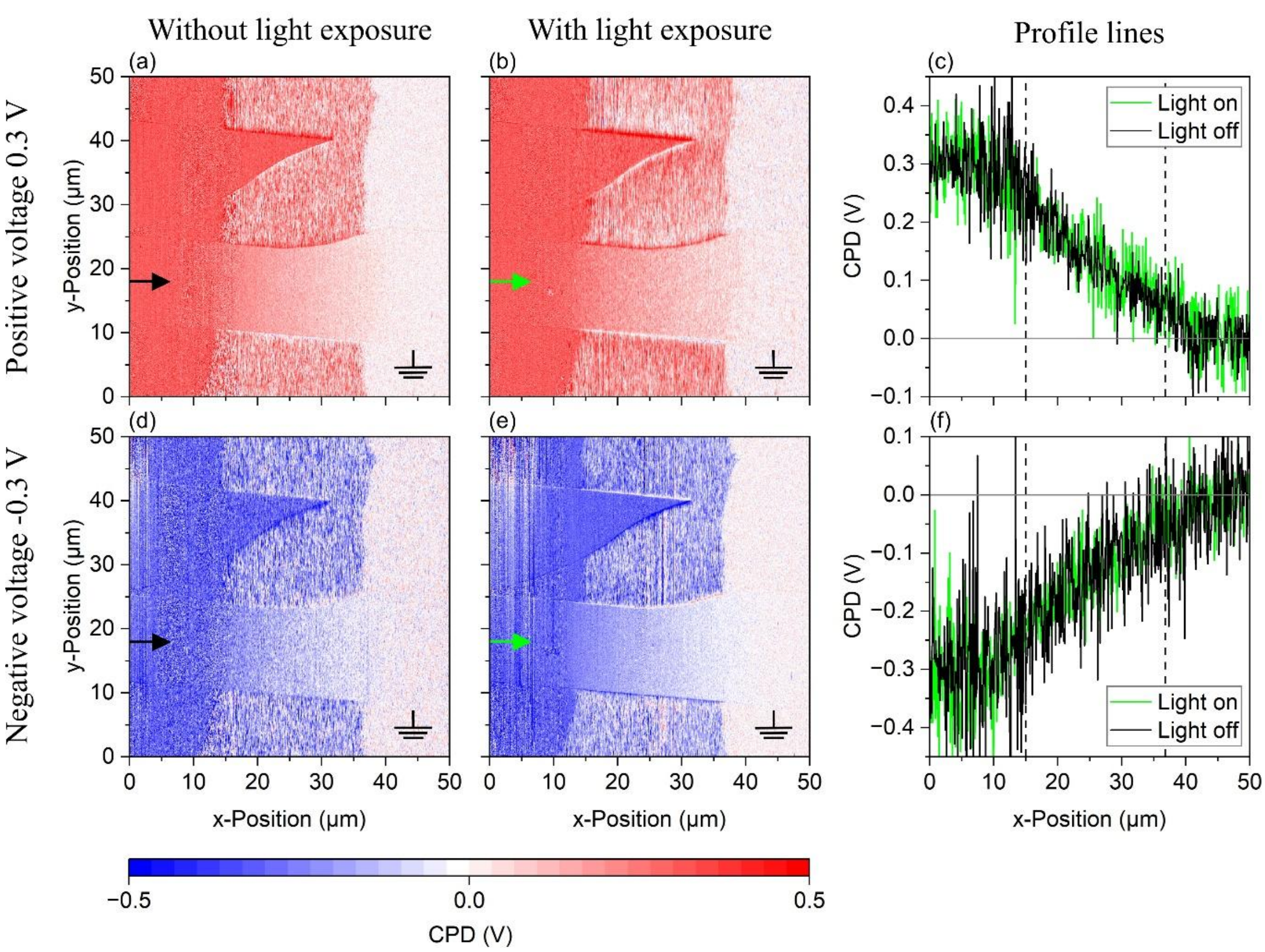

*Figure 10* The results of the KPFM measurements (in sideband mode under ambient conditions) are presented for sample O1. (a), (b) and (c) show the results of the measurements at a positive bias voltage of 0.3V and (d),(e) and (f) show the results of the measurements at a negative bias voltage of −0.3V. The right contact is set to ground potential whereby the positive or negative potential is applied to the left contact. The heat maps (a), (d) (without light exposure), (b) and (e) (with light exposure) present the lateral spatial dimensions in the x- and y-axis and the color code present the measured surface voltage (CPD at the sample bias voltage corrected by the CPD data) at a sample bias of 0V. Thus, the corrected data shows the effective potential changes on the sample induced by the applied voltage resulting in a potential distribution of the applied voltage in the contact system. In (c) and (f) the profile lines of the corrected CPD as a function of the x-coordinate are depicted. The black (green) function shows the profile line of the measurement without (with) light exposure. The y-coordinate at which the potentials are extracted for the profile lines are indicated by an arrow in the corresponding color (black or green) in the heat maps. The dashed lines demonstrate the contact edges. Under light exposure the sample was excited by laser light with an energy of $E_{laser}$ = 1.95 eV, that is above the band gap of $MoS_2$. The measurements with and without light exposure show no significant differences. A linear decrease of the potential across the flake without any abrupt potential drop is observed. This leads to the conclusion of very small and symmetric contact barriers.

## 4.2 KPFM potential data of Sample D1

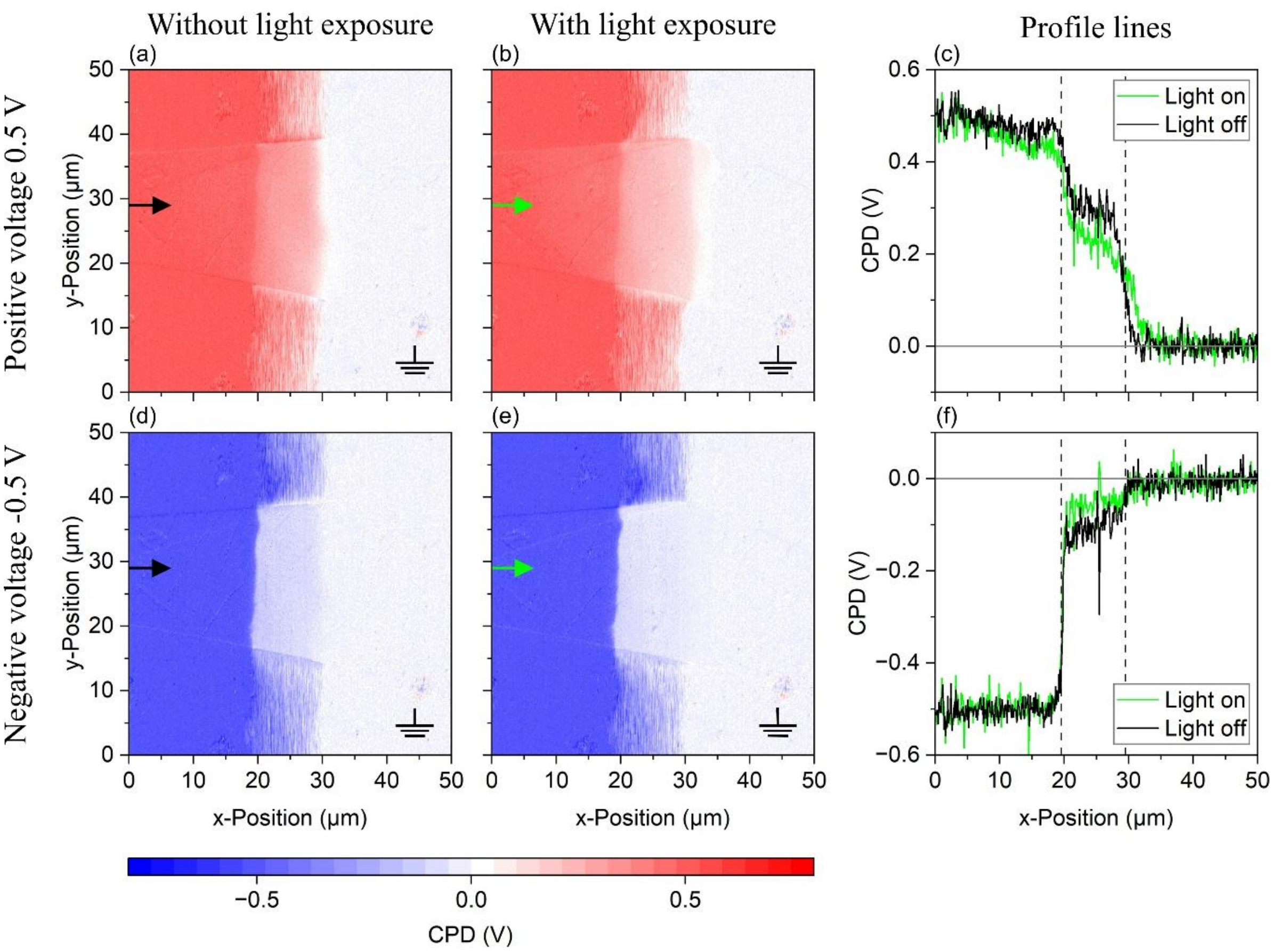

*Figure 11* The results of the KPFM measurements (in sideband mode under ambient conditions) are presented for sample D1. (a),(b) and (c) show the results of the measurements at a positive bias voltage of 0.5V and (d),(e) and (f) show the results of the measurements at a negative bias voltage of −0.5V. The right contact is set to ground potential whereby the positive or negative potential is applied to the left contact. The heat maps (a), (d) (without light exposure), (b) and (e) (with light exposure) present the lateral spatial dimensions in the x- and y-axis and the color code present the measured surface voltage (CPD at the sample bias voltage corrected by the CPD data) at a sample bias of 0V. Thus, the corrected data shows the effective potential changes on the sample induced by the applied voltage resulting in a potential distribution of the applied voltage in the contact system. In (c) and (f) the profile lines of the corrected CPD as a function of the x-coordinate are depicted. The black (green) function shows the profile line of the measurement without (with) light exposure. The y-coordinate at which the potentials are extracted for the profile lines are indicated by an arrow in the corresponding color (black or green) in the heat maps. The dashed lines demonstrate the contact edges. Under light exposure the sample was excited by laser light with an energy of $E_{laser}$ = 1.95 eV, that is above the band gap of $MoS_2$. A dominating and abrupt potential drop is observed at the contact edges. The potential drop at the two contact edges is unevenly distributed. The potential distribution depends on light exposure and on the voltage polarity. This leads to the conclusion of large and asymmetric contact barriers with a behavior like Schottky contacts.

## 4.3 Equivalent circuit of the contact system Au-$MoS_2$-Au

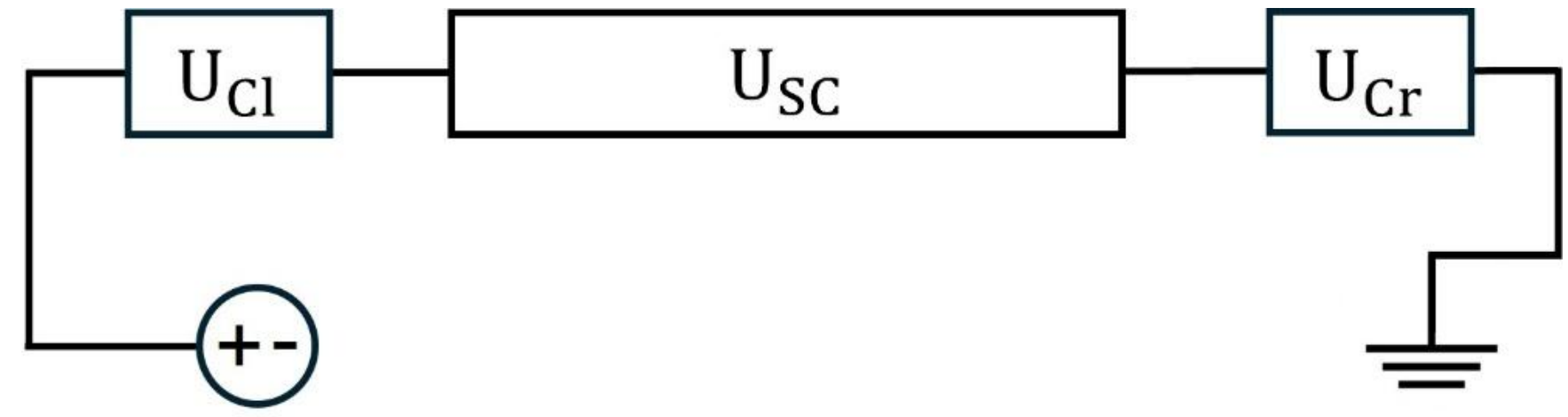

*Figure 12* Equivalent circuit of the contact system Au-$MoS_2$-Au consisting of three main resistors where the applied voltage is dropping relative to their relative resistivity for a given polarity: The equivalent resistance describing the left contact, the equivalent resistance of the semiconducting $MoS_2$ in the gap and the equivalent resistance describing the right contact. Since not the resistivity, but the relative voltage drop is measured, the circuit elements are marked by $U_{Cl}$, $U_{SC}$ and $U_{Cr}$, respectively. It has to be noted that this equivalent circuit is only valid at a static voltage. The resistors are voltage and light dependent and can contain contributions of a Schottky diode, a tunnel barrier, a photothermoelectric voltage and the resistance of the semiconductor.

## 4.4 Summary of the results of the KPFM measurements with an applied voltage of Sample O1 and D1:

| Sample O1 | Sample D1 |
|---|---|
| Approximately linear potential drops across the contact system | Abrupt potential drop at each contact edge |
| Potential distribution is independent of light exposure and the voltage polarity | Potential distribution is dependent on light exposure and the voltage polarity |
| Main potential drop across the flake within the gap | Main potential drop at the contact barriers |
| Very small and symmetric contact barriers → ohmic contacts, additional very small tunnel barriers possible | Large and asymmetric contact barriers → Schottky barriers potentially in combination with tunnel barriers |

## 4.5 KPFM potential data of sample D2 after annealing

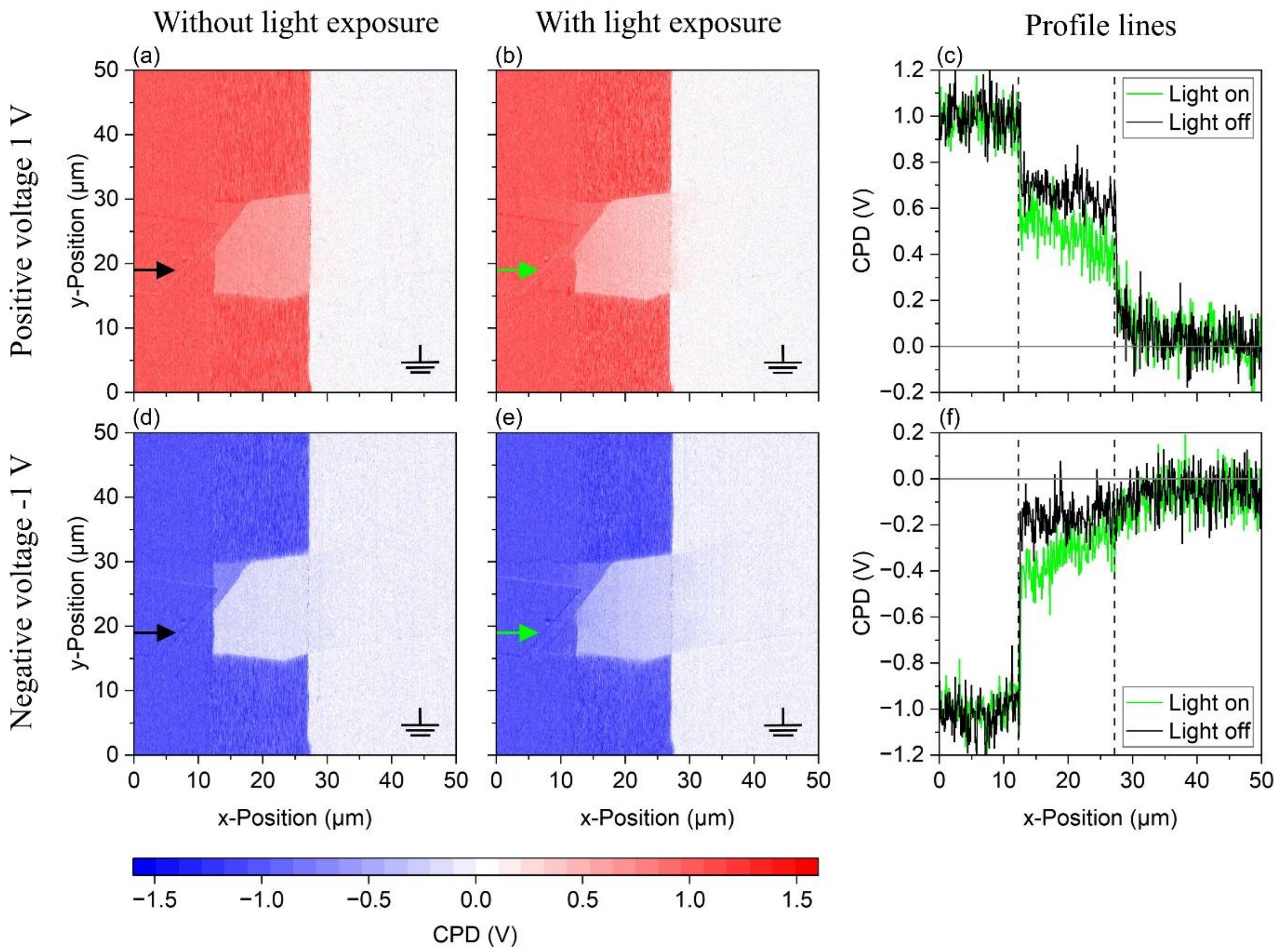

*Figure 13* The results of the KPFM measurements (in sideband mode under ambient conditions) are presented for sample D2. (a),(b) and (c) show the results of the measurements at a positive voltage of 1V and (d),(e) and (f) show the results of the measurements at a negative voltage of −1V. The right contact is set to ground potential whereby the positive or negative potential is applied to the left contact. The heat maps (a), (d) (without light exposure), (b) and (e) (with light exposure) present the lateral spatial dimensions in the x- and y-axis and the color code present the measured surface voltage (CPD at the sample bias voltage corrected by the CPD data) at a sample bias of 0V. Thus, the corrected data shows the effective potential changes on the sample induced by the applied voltage resulting in a potential distribution of the applied voltage in the contact system. In (c) and (f) the profile lines of the corrected CPD as a function of the x-coordinate are depicted. The black (green) function shows the profile line of the measurement without (with) light exposure. The y-coordinate at which the potentials are extracted for the profile lines are indicated by an arrow in the corresponding color (black or green) in the heat maps. The dashed lines demonstrate the contact edges. Under light exposure the sample was excited by laser light with an energy of $E_{laser}$ = 1.95 eV, that is above the band gap of $MoS_2$. A dominating and abrupt potential drop is observed at the contact edges. The potential drop at the two contact edges is unevenly distributed. The potential distribution depends on light exposure and on the voltage polarity. This leads to the conclusion of large and asymmetric contact barriers with a behavior like Schottky contacts similar to sample D1.

**KPFM potential profile lines at different positions**

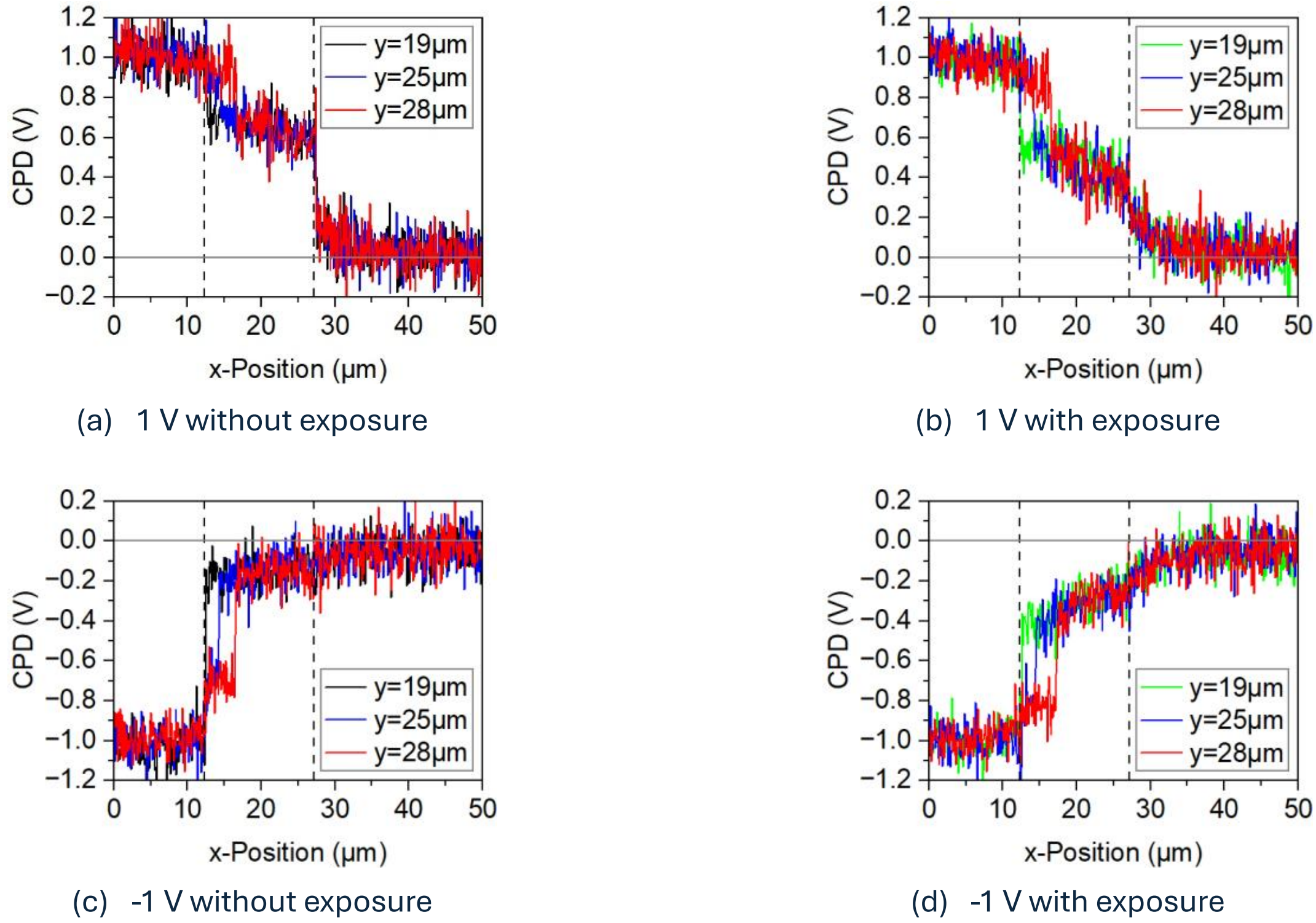

*Figure 14* Profile lines of the (corrected relative to 0V) KPFM potential of Sample D2 at different y-values (a) and (b) at the positive voltage and (c) and (d) at the negative voltage. The profile lines are shown with ((a) and (c)) and without ((b) and (d)) light exposure. The y-values 25 and 28 µm are in the upper half of the flake and cross the triangular area of the flake which shows a different behavior than the rest of the flake. The profile line at y=19 µm presents the data in the lower half of the flake without influences of the triangular area. These profile lines are the same lines that are already shown before. A shift of the x-position of the potential drop is observed which depends on the y-position where the profile line is taken. It is concluded, that this observation stems from a macroscopic distortion in the flake. As this distortion leads to a relevant voltage drop, it can be assumed that it represents a (tunnel) barrier at this position.

**Summary of the results of the KPFM measurements with an applied voltage of Sample D2 after annealing:**

- Abrupt voltage drop at each contact edge
- The potential distribution is dependent of light exposure and of the voltage polarity
- Main voltage drop at the contact barriers
- Large and asymmetric contact barriers
  - → Schottky barrier potentially in coexistence with a tunnel barrier
- Very similar KPFM results to those of sample D1 with the exception of the opoosite behaviour under light exposure at negative bias voltage

# 5. Measurement methods

## 5.1 Experimental Setup of electric transport measurements

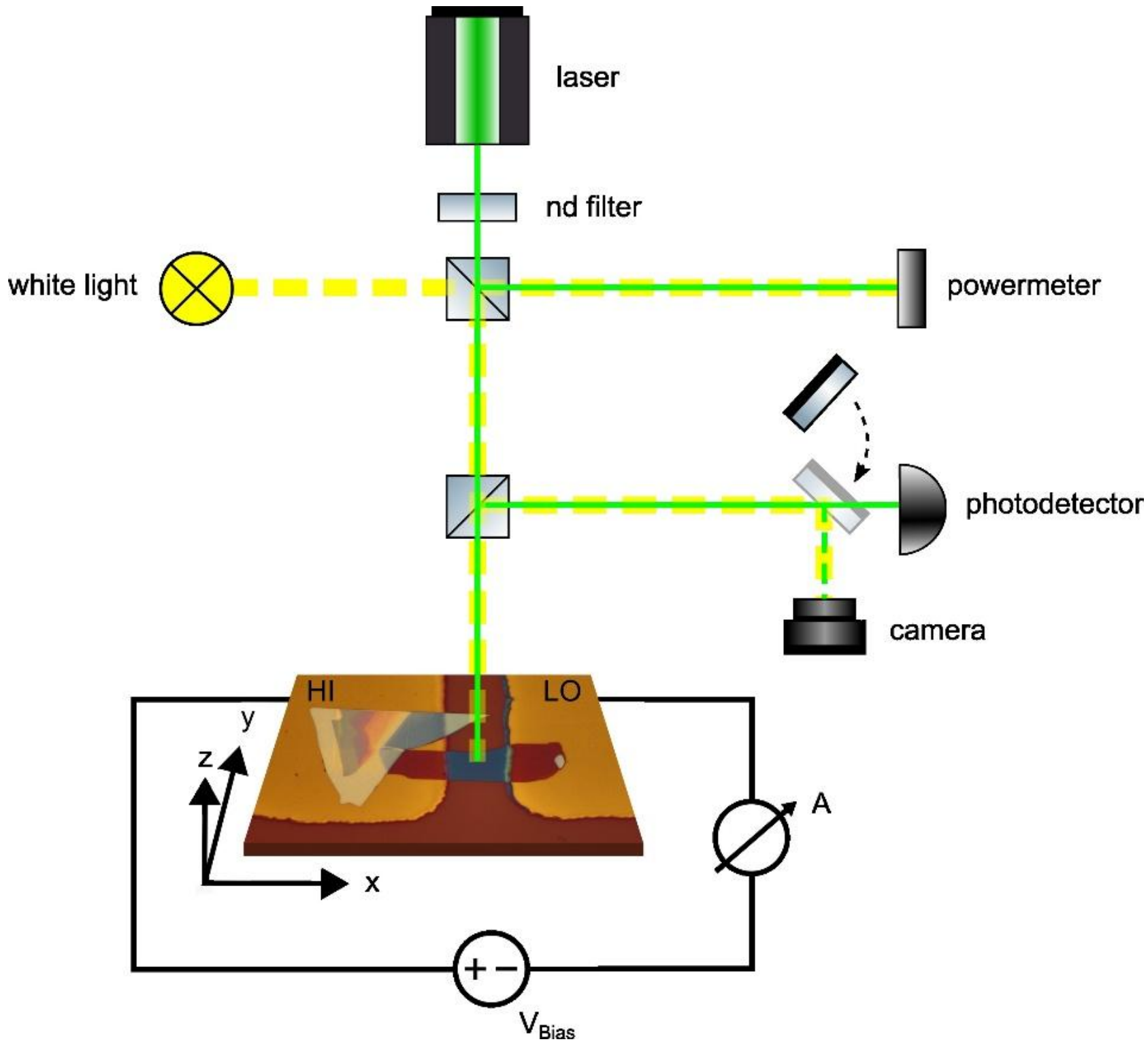

*Figure 15* Sketch of the setup of the IV and the local beam induced current (LBIC) measurements. A voltage is applied to the sample and the current is measured. The sample is excited by focused laser light (532 nm) of a specific laser power. The laser power is adjusted by the neutral density (nd) filter and measured by the powermeter. The z-position of the sample can be varied to adjust the size of the laser beam on the sample. For the IV measurements a defocused (global) excitation and for the photocurrent measurements a focused (local) excitation is applied. The sample can be moved motorized in x-and y-direction which is used for the scanning photocurrent measurement. The reflected light of the sample is collected by the objective and directed to a photodetector to measure the reflectance map of the sample. The white light source and the camera are used to image and align the laser beam on the sample. The sample is measured under ambient conditions.

## 5.2 Experimental Setup of Kelvin Probe Force Microscopy

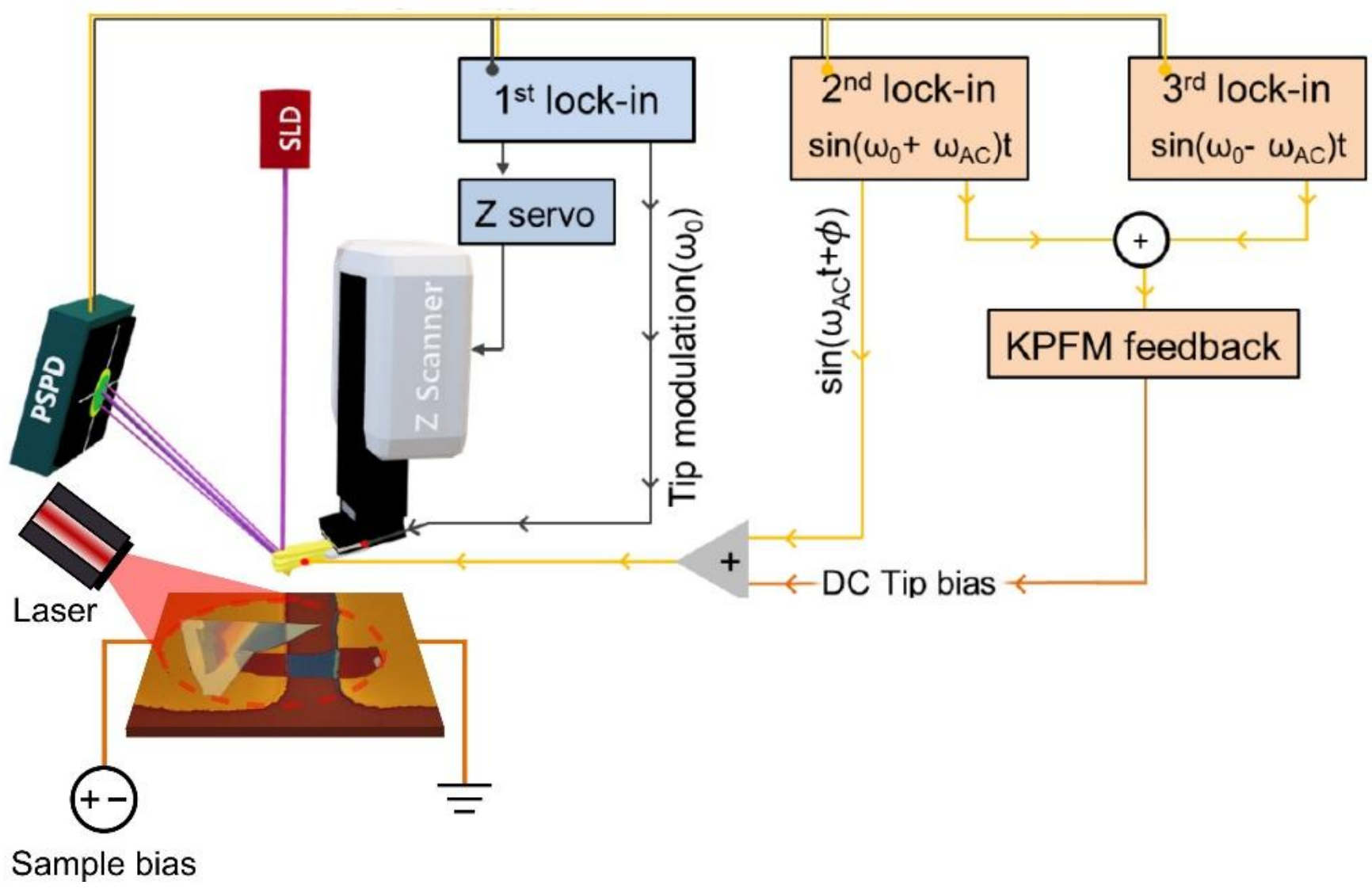

*Figure 16* Connection diagram of the Sideband-KPFM. A superluminescent diode (SLD) and a position-sensitive photodetector (PSPD) are used to detect the oscillation of the cantilever. An AC-voltage induces an oscillation of the tip above the sample. The first lock-in amplifier measures the frequency of the oscillation and the feedback-loop ensures a constant tip-sample-distance. Thus, the z-height of the sample is detected. Two additional lock-in amplifiers are used to measure the amplitude and phase of each sideband. This feedback loop regulates the DC voltage such that the sidebands disappear. The resulting DC voltage is the output signal and is equal to the CPD. A laser diode with the wavelength of 635 nm is used to illuminate the samples during the KPFM measurement. The sample is measured under ambient conditions.